\documentclass{cleiej}
\usepackage{graphicx}
\usepackage{hyperref}
\usepackage{doi}
\usepackage[utf8]{inputenc}
\usepackage{amsmath}
\usepackage{amsfonts}
\usepackage{amssymb}
\usepackage{lineno}
\usepackage{listings}
\usepackage{float}

\title{\texttt{rd-spiral}: An open-source Python library for learning 2D reaction-diffusion dynamics through pseudo-spectral method}

\author{
\bf Sandy H. S. Herho\\
Department of Earth and Planetary Sciences,\\ 
University of California, Riverside, CA, USA 92521, \\
\it sandy.herho@email.ucr.edu\\
\and
\bf Iwan P. Anwar \\
Bandung Institute of Technology (ITB),\\ 
Oceanography Research Group,\\ 
Bandung, West Java, Indonesia, 40132 \\
\it iwanpanwar@itb.ac.id 
\and
\bf Rusmawan Suwarman \\
Bandung Institute of Technology (ITB),\\
Atmospheric Science Research Group,\\ 
Bandung, West Java, Indonesia, 40132 \\
\it rusmawan@itb.ac.id
}
\begin{document}
\maketitle
\begin{abstract}
\noindent We present \texttt{rd-spiral}, an open-source Python library implementing pseudo-spectral methods for investigating two-dimensional reaction-diffusion dynamics with unprecedented pedagogical clarity and computational rigor. The software realizes a systematic mathematical framework derived from first-principles conservation laws, employing Fast Fourier Transform-based spatial discretization coupled with adaptive Dormand-Prince time integration to achieve exponential convergence while maintaining algorithmic transparency. Through comprehensive analysis of three distinct parameter regimes---stable spiral rotation, spatiotemporal chaos, and pattern decay---we demonstrate successful capture of the complete dynamical spectrum accessible within reaction-diffusion systems. Statistical characterization reveals extreme departures from Gaussian behavior in stable spiral dynamics, with leptokurtic distributions exhibiting excess kurtosis exceeding 96, necessitating development of a robust non-parametric analytical framework validated through 5,000 bootstrap iterations. Information-theoretic analysis quantifies the order-chaos transition through Shannon entropy evolution and mutual information decay, revealing a 10.7\% reduction in activator-inhibitor coupling during turbulent fragmentation compared to 6.5\% in stable regimes. The solver architecture demonstrates exceptional numerical efficiency, successfully handling stiffness ratios exceeding 6:1 while incorporating enhanced features including automated equilibrium classification, checkpoint mechanisms for extended simulations, and comprehensive error handling with partial result preservation. Effect size quantification using Cliff's delta reveals medium to large effects ($\delta = 0.37$--$0.78$) distinguishing dynamical regimes, with asymmetric field sensitivities suggesting differential responses to parameter perturbations. By prioritizing code clarity, comprehensive documentation, and reproducible parameter configurations, \texttt{rd-spiral} addresses the critical gap between theoretical formulations and practical implementation in nonlinear dynamics education. This work establishes computational protocols applicable to broader pattern-forming systems while demonstrating that pedagogically-oriented scientific software can achieve research-quality results without sacrificing the algorithmic transparency essential for educational advancement in computational physics and applied mathematics.
\end{abstract}


\section{Introduction}

The mathematical formalization of pattern formation in reaction-diffusion systems traces its origins to Turing's 1952 analysis of morphogenesis \cite{Turing1952}, which demonstrated that diffusion-driven instabilities could spontaneously break spatial symmetry in chemically reacting systems. This counterintuitive result---that diffusion, typically associated with homogenization, could generate heterogeneity---required precise conditions: differential diffusion rates between morphogens and specific reaction kinetics satisfying linear stability criteria. The theoretical framework remained purely mathematical for nearly four decades, lacking experimental validation and computational verification.

The experimental landscape shifted dramatically with the Belousov-Zhabotinsky (BZ) reaction, discovered accidentally by Belousov in 1959 while attempting to model the Krebs cycle \cite{Belousov1959}. His observations of periodic color changes in a homogeneous solution contradicted the prevailing thermodynamic dogma that chemical reactions must proceed monotonically toward equilibrium. The Soviet scientific establishment rejected his findings as impossible, delaying publication and recognition. Zhabotinsky's subsequent work \cite{Zhabotinsky1964} provided mechanistic understanding and reproducible protocols, establishing the BZ reaction as the canonical example of nonlinear chemical dynamics. These reactions exhibited not only temporal oscillations but also spatial patterns: target waves, rotating spirals, and spatiotemporal chaos.

The connection between abstract mathematical theory and chemical reality emerged through systematic analysis of excitable media. Winfree's 1972 topological approach to spiral waves \cite{Winfree1972} recognized their fundamental nature as phase singularities in two-dimensional oscillatory systems. His work established that spiral waves were not chemical curiosities but universal phenomena in excitable media, appearing across scales from intracellular calcium dynamics to cardiac tissue. The mathematical framework for analyzing these structures drew from diverse fields: differential geometry for spiral tip trajectories, bifurcation theory for stability analysis, and singular perturbation theory for multi-scale dynamics.

Neural excitation provided another crucial thread in this development. The Hodgkin-Huxley equations, while biophysically accurate, proved computationally intractable for large-scale simulations. FitzHugh's 1961 simplification \cite{FitzHugh1961} and Nagumo's circuit implementation \cite{Nagumo1962} reduced the four-variable ionic model to a two-variable system capturing essential excitability features. This FitzHugh-Nagumo model became the prototype for excitable reaction-diffusion systems, balancing mathematical tractability with biological relevance. The model's phase plane analysis revealed the geometric origins of excitability: a cubic nullcline intersecting a linear nullcline created the necessary slow-fast dynamics.

The rigorous mathematical treatment of reaction-diffusion waves evolved through the 1970s and 1980s. Kopell and Howard's 1973 analysis \cite{Kopell1973} established existence and stability conditions for plane waves in general reaction-diffusion systems. Keener's 1988 work on three-dimensional scroll waves \cite{Keener1988} extended the theory to higher dimensions, revealing new instabilities and complex dynamics. The medical relevance became apparent through the cardiac arrhythmia connection, first proposed by Wiener and Rosenblueth \cite{Wiener1946} and developed extensively by Winfree \cite{Winfree1987}. Spiral waves in cardiac tissue could degenerate into turbulent patterns resembling ventricular fibrillation, suggesting therapeutic targets for defibrillation strategies.

Computational methods for reaction-diffusion systems evolved in parallel with theoretical understanding. Early simulations by Hagan \cite{Hagan1982} used finite differences on coarse grids, limited by computational resources to qualitative observations. The introduction of spectral methods by Gottlieb and Orszag \cite{Gottlieb1977} offered exponential convergence for smooth solutions, but their application to pattern-forming systems required careful treatment. The periodic boundary conditions natural to Fourier methods matched the mathematical requirements for studying bulk pattern properties, while the global nature of spectral differentiation captured long-range interactions more accurately than local finite difference stencils.

The stiffness inherent in reaction-diffusion systems---arising from disparate time scales between fast reactions and slow diffusion---demanded specialized numerical methods. Traditional explicit schemes suffered from severe timestep restrictions, while fully implicit methods required solving large nonlinear systems at each step. Cox and Matthews \cite{Cox2002} developed exponential time differencing (ETD) methods that treated the linear diffusion exactly through matrix exponentials while handling nonlinear reactions explicitly. Kassam and Trefethen's fourth-order implementation \cite{Kassam2005} combined ETD with Runge-Kutta methods, achieving both stability and accuracy for stiff pattern-forming systems.

Experimental validation of Turing patterns proved elusive until Castets et al.\cite{Castets1990} demonstrated stationary patterns in the chlorite-iodide-malonic acid (CIMA) reaction. Their success required precise control of reaction conditions and gel reactors to suppress convection. This experimental breakthrough validated Turing's theoretical predictions and sparked renewed interest in biological pattern formation. Subsequent work identified Turing-type mechanisms in diverse biological systems: pigmentation patterns, digit formation, and cortical development.


The computational infrastructure for scientific computing underwent radical transformation between 1990 and 2020. MATLAB\textsuperscript{\textregistered}'s introduction democratized numerical computing but imposed licensing barriers and proprietary constraints. The emergence of Python as a scientific computing platform, driven by NumPy \cite{Harris2020} and SciPy \cite{Virtanen2020}, created an open ecosystem accessible globally. Recent comparative studies have demonstrated Python's superior performance in various scientific computing applications: Python achieved 63.57\% faster execution times than R in spring-mass-damper control system simulations \cite{herho2025quantitative}, outperformed Julia by 18.6$\times$ in Kalman filter implementations despite theoretical expectations \cite{kaban2024performance}, and showed competitive performance against Fortran, Julia, and R in 2D Lagrangian fluid dynamics simulations \cite{herho2024comparing}. 

The pedagogical challenges in teaching reaction-diffusion dynamics stem from their inherent interdisciplinarity. Students must integrate concepts from partial differential equations, numerical analysis, nonlinear dynamics, and physical chemistry. Traditional approaches compartmentalize these topics, leaving students to synthesize connections independently. Strogatz \cite{Strogatz2015} emphasized computational experiments as essential for developing intuition about nonlinear systems, yet practical implementation remained challenging. Research codes, optimized for performance and generality, obscure the underlying algorithms through layers of abstraction and optimization.

The reproducibility crisis in computational science, documented by Wilson et al.\cite{Wilson2014}, highlighted systematic failures in code sharing and documentation. Published papers often omitted crucial implementation details, making reproduction impossible even with access to source code. The proliferation of ad hoc scripts and undocumented research codes created barriers to verification and extension. These challenges were particularly acute in reaction-diffusion modeling, where subtle numerical choices---boundary conditions, time integration schemes, spatial discretization---could qualitatively alter results.

Current software for reaction-diffusion modeling reflects tensions between generality, performance, and accessibility. Finite element packages like FEniCS \cite{Logg2012} and DUNE \cite{Bastian2021} provide sophisticated spatial discretization but require expertise in variational formulations and mesh generation. Specialized biological simulators like Virtual Cell \cite{Moraru2008} and COPASI \cite{Hoops2006} offer user-friendly interfaces but hide numerical methods behind black-box implementations. General-purpose tools in MATLAB\textsuperscript{\textregistered} and Wolfram Mathematica provide middle ground but require commercial licenses and offer limited customization.

The machine learning revolution has renewed interest in partial differential equations through physics-informed neural networks (PINNs) \cite{Raissi2019}. These methods promise to bypass traditional numerical discretization but require validation against established techniques. The integration of machine learning with classical numerical methods demands deep understanding of both domains. Students trained exclusively in data-driven methods lack appreciation for conservation laws, stability constraints, and asymptotic behavior that traditional methods encode naturally.

The COVID-19 pandemic accelerated computational methods adoption across scientific disciplines. Experimentalists pivoted to modeling when laboratory access was restricted, creating urgent demand for accessible computational tools. Remote education highlighted the importance of self-contained, well-documented software that students could run independently. The shift toward open science practices, exemplified by preprint servers and code repositories, created new venues for sharing educational resources.

Spiral waves in reaction-diffusion systems offer an ideal pedagogical entry point connecting abstract mathematics to observable phenomena. These structures appear across biological scales: intracellular calcium waves \cite{Lechleiter1991}, cardiac electrical activity \cite{Davidenko1992}, and population dynamics \cite{Hassell1991}. Their mathematical properties---topological charge conservation, parameter-dependent stability, bifurcations to turbulence---illustrate fundamental concepts in nonlinear dynamics. Computational generation of spiral waves requires understanding initial conditions, boundary effects, and numerical stability, providing concrete examples of abstract numerical concepts.

The pseudo-spectral approach implemented in \texttt{rd-spiral} leverages the Fast Fourier Transform (FFT)'s ubiquity across computational science. Students encounter FFTs in signal processing, image analysis, and machine learning, but rarely appreciate their application to differential equations. The method's elegance---transforming differential operators to algebraic multiplication---provides insight into the deep connection between functional analysis and numerical computation. The periodic boundary conditions inherent in Fourier methods naturally suit pattern formation studies where bulk properties matter more than edge effects.

Our implementation philosophy prioritizes transparency over performance, recognizing that premature optimization obscures algorithmic understanding. Each computational step---forward transform, nonlinear evaluation, inverse transform, time integration---corresponds directly to the mathematical formulation. The modular structure allows component modification without system-wide changes, facilitating experimentation. By avoiding advanced optimizations like operator splitting or adaptive grids, we maintain conceptual clarity while achieving research-quality results.

The parameter regimes explored in \texttt{rd-spiral}---stable spirals, turbulent dynamics, pattern decay---represent distinct dynamical behaviors accessible through simple parameter changes. This allows systematic exploration of bifurcations without extensive parameter searches. The pre-configured examples reliably produce expected behaviors, avoiding the frustration of failed simulations that plague initial computational experiments. The visual nature of pattern formation provides immediate feedback, allowing students to develop intuition through observation before engaging with mathematical analysis.

The broader context for this work encompasses the democratization of computational science through open-source tools and the recognition of software as scholarly contribution. Contemporary scientific progress increasingly depends on accessible computational implementations that enable researchers to build upon existing work, verify results, and extend methodologies. The establishment of rigorous peer review standards for research software reflects a fundamental shift in how the scientific community values documentation, testing, and community impact alongside algorithmic novelty. Our contribution aligns with these evolving principles by providing transparent, well-documented code that bridges the gap between theoretical formulations and practical implementation. By prioritizing pedagogical clarity and reproducibility, \texttt{rd-spiral} serves dual purposes: as an educational tool for students learning numerical methods and as a reliable foundation for researchers investigating reaction-diffusion dynamics. This approach acknowledges that scientific advancement requires not only theoretical breakthroughs but also the computational infrastructure that makes such theories testable and extensible by the broader scientific community.

\section{Methods}
\subsection{Mathematical Model}

The mathematical formulation of reaction-diffusion systems requires a systematic derivation from first principles to establish both the governing equations and the physical interpretation of each term. We begin with the most fundamental conservation laws and progressively build toward the specific model that exhibits spiral wave dynamics. This comprehensive derivation not only establishes mathematical rigor but also provides deep insight into the physical mechanisms and approximations inherent in the model.

Consider a two-dimensional domain $\Omega \subset \mathbb{R}^2$ containing a reactive chemical medium. Within this domain, we track the evolution of two chemical species with molar concentrations $U(\mathbf{x},t)$ and $V(\mathbf{x},t)$, where $\mathbf{x} = (X,Y)$ represents the spatial position vector and $t \in [0,\infty)$ denotes time. The fundamental principle governing any conserved quantity in a continuum is that the rate of change of the total amount within any control volume must equal the net flux through its boundary plus any internal sources or sinks \cite{Reddy2017,Kleiber2016}. For an arbitrary control volume $\mathcal{V} \subset \Omega$ with boundary $\partial\mathcal{V}$, the integral form of mass conservation for species $i$ states:
\begin{equation}
\frac{d}{dt}\int_{\mathcal{V}} C_i(\mathbf{x},t) \, dV = -\oint_{\partial\mathcal{V}} \mathbf{J}_i(\mathbf{x},t) \cdot \mathbf{n} \, dS + \int_{\mathcal{V}} S_i(\mathbf{x},t) \, dV,
\label{eq:integral_conservation_general}
\end{equation}
where $C_i$ represents the concentration field of species $i \in \{U,V\}$, $\mathbf{J}_i$ denotes the mass flux vector (amount per unit area per unit time), $\mathbf{n}$ is the outward-pointing unit normal vector to the surface element $dS$, and $S_i$ represents the net rate of production of species $i$ due to chemical reactions.

The surface integral in equation \eqref{eq:integral_conservation_general} represents the total flux of mass through the boundary. To transform this into a more tractable form, we invoke the divergence theorem (also known as Gauss's theorem), which provides a fundamental connection between surface integrals and volume integrals. For any continuously differentiable vector field $\mathbf{F}$ defined on a region $\mathcal{V}$ with piecewise smooth boundary $\partial\mathcal{V}$:
\begin{equation}
\oint_{\partial\mathcal{V}} \mathbf{F} \cdot \mathbf{n} \, dS = \int_{\mathcal{V}} \nabla \cdot \mathbf{F} \, dV,
\label{eq:divergence_theorem_statement}
\end{equation}
where the divergence operator in Cartesian coordinates is defined as:
\begin{equation}
\nabla \cdot \mathbf{F} \equiv \frac{\partial F_x}{\partial X} + \frac{\partial F_y}{\partial Y},
\label{eq:divergence_definition}
\end{equation}
with $F_x$ and $F_y$ being the $X$ and $Y$ components of the vector field $\mathbf{F}$.

Applying the divergence theorem to the flux term in equation \eqref{eq:integral_conservation_general}:
\begin{equation}
\oint_{\partial\mathcal{V}} \mathbf{J}_i \cdot \mathbf{n} \, dS = \int_{\mathcal{V}} \nabla \cdot \mathbf{J}_i \, dV.
\label{eq:divergence_applied_flux}
\end{equation}
Substituting this result back into the conservation equation:
\begin{equation}
\frac{d}{dt}\int_{\mathcal{V}} C_i \, dV = -\int_{\mathcal{V}} \nabla \cdot \mathbf{J}_i \, dV + \int_{\mathcal{V}} S_i \, dV.
\label{eq:conservation_divergence_form}
\end{equation}

For a fixed control volume (not moving with time), we can interchange the time derivative with the spatial integration:
\begin{equation}
\frac{d}{dt}\int_{\mathcal{V}} C_i \, dV = \int_{\mathcal{V}} \frac{\partial C_i}{\partial t} \, dV.
\label{eq:leibniz_rule}
\end{equation}
This application of Leibniz's rule for differentiation under the integral sign yields:
\begin{equation}
\int_{\mathcal{V}} \frac{\partial C_i}{\partial t} \, dV = -\int_{\mathcal{V}} \nabla \cdot \mathbf{J}_i \, dV + \int_{\mathcal{V}} S_i \, dV.
\label{eq:integral_combined}
\end{equation}

Rearranging all terms under a single integral:
\begin{equation}
\int_{\mathcal{V}} \left[\frac{\partial C_i}{\partial t} + \nabla \cdot \mathbf{J}_i - S_i\right] dV = 0.
\label{eq:integral_zero}
\end{equation}
The crucial observation is that equation \eqref{eq:integral_zero} must hold for any arbitrary choice of control volume $\mathcal{V}$ within our domain. If the integrand were non-zero at some point $\mathbf{x}_0$, we could construct a small control volume around that point where the integral would be non-zero, contradicting equation \eqref{eq:integral_zero}. Therefore, the integrand must vanish identically at every point:
\begin{equation}
\frac{\partial C_i}{\partial t} + \nabla \cdot \mathbf{J}_i - S_i = 0 \quad \text{for all } \mathbf{x} \in \Omega.
\label{eq:pointwise_conservation}
\end{equation}

The mass flux $\mathbf{J}_i$ requires a constitutive relationship to close the system. For molecular diffusion in dilute solutions, extensive experimental evidence supports Fick's law \cite{laBarrera2005,Poirier2016}, which postulates that the flux is proportional to the negative concentration gradient. In its most general tensorial form:
\begin{equation}
\mathbf{J}_i = -\mathcal{D}_i \cdot \nabla C_i,
\label{eq:ficks_tensor}
\end{equation}
where $\mathcal{D}_i$ is the diffusion tensor. For isotropic media, where diffusion properties are independent of direction, the tensor reduces to a scalar multiple of the identity:
\begin{equation}
\mathcal{D}_i = \mathcal{D}_i \mathbf{I} = \mathcal{D}_i \begin{pmatrix} 1 & 0 \\ 0 & 1 \end{pmatrix},
\label{eq:isotropic_diffusion}
\end{equation}
yielding the familiar form:
\begin{equation}
\mathbf{J}_i = -\mathcal{D}_i \nabla C_i = -\mathcal{D}_i \left(\frac{\partial C_i}{\partial X}\mathbf{e}_x + \frac{\partial C_i}{\partial Y}\mathbf{e}_y\right),
\label{eq:ficks_isotropic_expanded}
\end{equation}
where $\mathbf{e}_x$ and $\mathbf{e}_y$ are the unit vectors in the $X$ and $Y$ directions.

To compute the divergence of the flux, we must consider whether the diffusion coefficient varies spatially. In the general case:
\begin{equation}
\nabla \cdot \mathbf{J}_i = -\nabla \cdot (\mathcal{D}_i \nabla C_i) = -\frac{\partial}{\partial X}\left(\mathcal{D}_i \frac{\partial C_i}{\partial X}\right) - \frac{\partial}{\partial Y}\left(\mathcal{D}_i \frac{\partial C_i}{\partial Y}\right).
\label{eq:divergence_general}
\end{equation}

Applying the product rule to each term:
\begin{align}
\frac{\partial}{\partial X}\left(\mathcal{D}_i \frac{\partial C_i}{\partial X}\right) &= \frac{\partial \mathcal{D}_i}{\partial X}\frac{\partial C_i}{\partial X} + \mathcal{D}_i\frac{\partial^2 C_i}{\partial X^2}, \label{eq:product_rule_x}\\
\frac{\partial}{\partial Y}\left(\mathcal{D}_i \frac{\partial C_i}{\partial Y}\right) &= \frac{\partial \mathcal{D}_i}{\partial Y}\frac{\partial C_i}{\partial Y} + \mathcal{D}_i\frac{\partial^2 C_i}{\partial Y^2}. \label{eq:product_rule_y}
\end{align}

For homogeneous media where the diffusion coefficient is spatially constant, the gradient terms vanish:
\begin{equation}
\frac{\partial \mathcal{D}_i}{\partial X} = \frac{\partial \mathcal{D}_i}{\partial Y} = 0,
\label{eq:constant_diffusion_condition}
\end{equation}
simplifying the divergence to:
\begin{equation}
\nabla \cdot \mathbf{J}_i = -\mathcal{D}_i \left(\frac{\partial^2 C_i}{\partial X^2} + \frac{\partial^2 C_i}{\partial Y^2}\right) \equiv -\mathcal{D}_i \nabla^2 C_i,
\label{eq:laplacian_flux}
\end{equation}
where $\nabla^2$ is the Laplacian operator.

Substituting this result into equation \eqref{eq:pointwise_conservation} yields the reaction-diffusion equation:
\begin{equation}
\frac{\partial C_i}{\partial t} = \mathcal{D}_i \nabla^2 C_i + S_i(C_U, C_V, \mathbf{x}, t).
\label{eq:reaction_diffusion_general}
\end{equation}

The source term $S_i$ encodes the chemical kinetics. To derive its functional form systematically, we consider the system near a reference state and expand in a Taylor series \cite{Albagli1990}. Let $(U_0, V_0)$ represent a spatially uniform reference concentration (not necessarily an equilibrium). Expanding $S_U$ about this reference:
\begin{align}
S_U(U,V) &= S_U(U_0,V_0) + \frac{\partial S_U}{\partial U}\bigg|_{U_0,V_0}(U-U_0) + \frac{\partial S_U}{\partial V}\bigg|_{U_0,V_0}(V-V_0) \nonumber\\
&\quad + \frac{1}{2}\frac{\partial^2 S_U}{\partial U^2}\bigg|_{U_0,V_0}(U-U_0)^2 + \frac{\partial^2 S_U}{\partial U \partial V}\bigg|_{U_0,V_0}(U-U_0)(V-V_0) \nonumber\\
&\quad + \frac{1}{2}\frac{\partial^2 S_U}{\partial V^2}\bigg|_{U_0,V_0}(V-V_0)^2 + \text{higher order terms}.
\label{eq:taylor_expansion_full}
\end{align}

For pattern-forming systems, we require that the reference state be unstable to small perturbations. This typically occurs when the linear terms promote growth. Additionally, to prevent unbounded growth, nonlinear terms must provide saturation. The minimal model retaining these features includes linear and cubic terms. Shifting the origin to $(U_0, V_0) = (0,0)$ (which can always be done by redefining variables) and keeping only odd-order terms (which ensures the symmetry $(U,V) \rightarrow -(U,V)$ preserves the form of equations):
\begin{equation}
S_U(U,V) = a_{10}U + a_{12}UV^2 + a_{30}U^3 + a_{21}U^2V + a_{03}V^3 + O(|(U,V)|^5),
\label{eq:taylor_truncated_u}
\end{equation}
where the coefficients $a_{jk}$ represent derivatives:
\begin{equation}
a_{jk} = \frac{1}{j!k!}\frac{\partial^{j+k} S_U}{\partial U^j \partial V^k}\bigg|_{(0,0)}.
\label{eq:taylor_coefficients}
\end{equation}

Physical and mathematical considerations constrain these coefficients. The linear term $a_{10}$ must be positive for instability. The cubic self-interaction $a_{30}$ must be negative to prevent unbounded growth. Cross-interactions can have either sign, but their relative magnitudes determine the dynamical behavior. Writing these coefficients in terms of rate constants with appropriate dimensions:
\begin{equation}
S_U = k_1 U - k_2 U^3 - k_3 UV^2 + k_4 U^2V + k_5 V^3,
\label{eq:reaction_u_dimensional}
\end{equation}
where $k_1$ has dimension $[\text{time}^{-1}]$, while $k_2$ through $k_5$ have dimension $[\text{concentration}^{-2}\text{time}^{-1}]$.

Similarly, for the second species:
\begin{equation}
S_V = k_6 V - k_7 U^2V - k_8 V^3 - k_9 U^3 - k_{10} UV^2.
\label{eq:reaction_v_dimensional}
\end{equation}

The system now contains twelve parameters (ten rate constants plus two diffusion coefficients), making systematic analysis challenging. Dimensional analysis provides a powerful tool for reducing this complexity \cite{Collis2016}. We introduce characteristic scales:
\begin{itemize}
\item Length scale $L_c$: typically the domain size or pattern wavelength
\item Time scale $T_c$: to be determined from the dynamics
\item Concentration scale $C_c$: to be determined from the kinetics
\end{itemize}

Define dimensionless variables:
\begin{equation}
x \equiv \frac{X}{L_c}, \quad y \equiv \frac{Y}{L_c}, \quad \tau \equiv \frac{t}{T_c}, \quad u \equiv \frac{U}{C_c}, \quad v \equiv \frac{V}{C_c}.
\label{eq:dimensionless_definitions}
\end{equation}

The partial derivatives transform according to the chain rule:
\begin{align}
\frac{\partial}{\partial t} &= \frac{\partial \tau}{\partial t}\frac{\partial}{\partial \tau} = \frac{1}{T_c}\frac{\partial}{\partial \tau}, \label{eq:time_derivative_transform}\\
\frac{\partial}{\partial X} &= \frac{\partial x}{\partial X}\frac{\partial}{\partial x} = \frac{1}{L_c}\frac{\partial}{\partial x}, \label{eq:space_derivative_transform}\\
\frac{\partial^2}{\partial X^2} &= \frac{1}{L_c^2}\frac{\partial^2}{\partial x^2}. \label{eq:second_derivative_transform}
\end{align}

Applying these transformations to the reaction-diffusion equation for species $U$:
\begin{equation}
\frac{C_c}{T_c}\frac{\partial u}{\partial \tau} = \mathcal{D}_U \frac{C_c}{L_c^2}\left(\frac{\partial^2 u}{\partial x^2} + \frac{\partial^2 u}{\partial y^2}\right) + S_U(C_c u, C_c v).
\label{eq:transformed_equation}
\end{equation}

Multiplying through by $T_c/C_c$:
\begin{equation}
\frac{\partial u}{\partial \tau} = \frac{\mathcal{D}_U T_c}{L_c^2}\nabla^2 u + \frac{T_c}{C_c}S_U(C_c u, C_c v),
\label{eq:nondim_intermediate_form}
\end{equation}
where $\nabla^2 \equiv \partial^2/\partial x^2 + \partial^2/\partial y^2$ is now the dimensionless Laplacian.

The natural choice for the time scale is $T_c = 1/k_1$, which normalizes the linear growth rate. This yields a dimensionless diffusion coefficient:
\begin{equation}
D_1 \equiv \frac{\mathcal{D}_U T_c}{L_c^2} = \frac{\mathcal{D}_U}{k_1 L_c^2}.
\label{eq:dimensionless_diffusion}
\end{equation}
This parameter represents the ratio of the diffusive time scale $L_c^2/\mathcal{D}_U$ to the reactive time scale $1/k_1$.

Expanding the reaction term:
\begin{align}
\frac{T_c}{C_c}S_U(C_c u, C_c v) &= \frac{1}{k_1 C_c}[k_1 C_c u - k_2 C_c^3 u^3 - k_3 C_c^3 uv^2 + k_4 C_c^3 u^2v + k_5 C_c^3 v^3] \nonumber\\
&= u - \frac{k_2 C_c^2}{k_1}u^3 - \frac{k_3 C_c^2}{k_1}uv^2 + \frac{k_4 C_c^2}{k_1}u^2v + \frac{k_5 C_c^2}{k_1}v^3.
\label{eq:reaction_expanded}
\end{align}

The concentration scale emerges naturally by requiring the coefficient of $u^3$ to be unity:
\begin{equation}
\frac{k_2 C_c^2}{k_1} = 1 \quad \Rightarrow \quad C_c = \sqrt{\frac{k_1}{k_2}}.
\label{eq:concentration_scale}
\end{equation}
This choice represents the concentration at which linear growth balances cubic saturation.

With this scaling, the dimensionless parameters become:
\begin{equation}
\alpha_3 \equiv \frac{k_3}{k_2}, \quad \beta_1 \equiv \frac{k_4}{k_2}, \quad \gamma \equiv \frac{k_5}{k_2}.
\label{eq:dimensionless_parameters}
\end{equation}

Following identical procedures for species $V$ and introducing symmetry constraints to reduce the parameter space, we arrive at the canonical form. Setting equal linear growth rates, symmetric saturation, and parameterizing the coupling through a single parameter $\beta$, the final dimensionless system reads \cite{Bramburger2024}:
\begin{equation}
\frac{\partial u}{\partial \tau} = D_1 \nabla^2 u + f(u,v), \quad f(u,v) \equiv u - u^3 - uv^2 + \beta(u^2v + v^3),
\label{eq:final_u_complete}
\end{equation}
\begin{equation}
\frac{\partial v}{\partial \tau} = D_2 \nabla^2 v + g(u,v), \quad g(u,v) \equiv v - u^2v - v^3 - \beta(u^3 + uv^2).
\label{eq:final_v_complete}
\end{equation}

This systematic derivation from conservation laws through dimensional analysis reveals that the essential dynamics depend on only three parameters: $D_1$, $D_2$, and $\beta$. The diffusion coefficients control the spatial coupling strength, while $\beta$ determines the nature of the nonlinear interactions. When $\beta = 0$, the system possesses a variational structure, while $\beta \neq 0$ enables the non-equilibrium dynamics essential for spiral wave formation.

\subsection{Numerical Implementation}

The theoretical formulation presented in the previous subsection requires careful numerical implementation to maintain accuracy while preserving pedagogical transparency. The foundation of our approach rests on the pseudo-spectral method, which we derive from first principles to establish the mathematical basis for the computational algorithm. Consider the reaction-diffusion system in \eqref{eq:final_u_complete}--\eqref{eq:final_v_complete} on a periodic domain $\Omega = [0, L]^2$. The periodicity assumption allows us to expand any sufficiently smooth function in a Fourier series. For a function $f(x,y)$ satisfying periodic boundary conditions $f(x+L,y) = f(x,y)$ and $f(x,y+L) = f(x,y)$, the continuous Fourier expansion takes the form:
\begin{equation}
f(x,y) = \sum_{k_x=-\infty}^{\infty} \sum_{k_y=-\infty}^{\infty} \hat{f}_{k_x,k_y} \exp\left(i\frac{2\pi}{L}(k_x x + k_y y)\right),
\label{eq:continuous_fourier_expansion}
\end{equation}
where the Fourier coefficients are given by:
\begin{equation}
\hat{f}_{k_x,k_y} = \frac{1}{L^2} \int_0^L \int_0^L f(x,y) \exp\left(-i\frac{2\pi}{L}(k_x x + k_y y)\right) dx dy.
\label{eq:fourier_coefficients}
\end{equation}

The orthogonality of the complex exponentials ensures that this representation is unique. Specifically, for integer wavenumbers $k$ and $m$:
\begin{equation}
\frac{1}{L} \int_0^L \exp\left(i\frac{2\pi}{L}(k-m)x\right) dx = \delta_{km},
\label{eq:orthogonality}
\end{equation}
where $\delta_{km}$ is the Kronecker delta. This orthogonality property is fundamental to the spectral accuracy of Fourier methods \cite{Boyd2001}.

For numerical computation, we must truncate the infinite series and discretize the spatial domain. Consider a uniform grid with $n$ points in each direction: $x_j = j\Delta x$ and $y_k = k\Delta y$ where $j,k = 0,1,\ldots,n-1$ and $\Delta x = \Delta y = L/n$. The discrete Fourier transform (DFT) approximates the continuous transform by replacing integrals with Riemann sums:
\begin{equation}
\hat{f}_{k_x,k_y} \approx \frac{1}{n^2} \sum_{j=0}^{n-1} \sum_{k=0}^{n-1} f_{jk} \exp\left(-i\frac{2\pi}{n}(k_x j + k_y k)\right),
\label{eq:dft_forward}
\end{equation}
where $f_{jk} = f(x_j, y_k)$ and the discrete wavenumbers are restricted to $k_x, k_y \in \{0,1,\ldots,n-1\}$. The inverse discrete transform recovers the spatial values:
\begin{equation}
f_{jk} = \sum_{k_x=0}^{n-1} \sum_{k_y=0}^{n-1} \hat{f}_{k_x,k_y} \exp\left(i\frac{2\pi}{n}(k_x j + k_y k)\right).
\label{eq:dft_inverse}
\end{equation}

The aliasing phenomenon arises from the finite sampling rate. By the Nyquist-Shannon theorem, we can only accurately represent frequencies up to $k_{\max} = n/2$. Higher frequencies alias to lower ones through the relation $k \equiv k \bmod n$. To maintain real-valued functions after inverse transformation, we adopt the standard convention of ordering wavenumbers as $k \in \{0,1,\ldots,n/2-1,-n/2,\ldots,-1\}$, which maps to physical wavenumbers $k_{\text{phys}} = 2\pi k/L$ \cite{Frigo2005}.

The FFT algorithm, developed by Cooley and Tukey, reduces the computational complexity from $O(n^4)$ for direct evaluation to $O(n^2 \log n)$ through recursive decomposition \cite{Cooley1965}. For a one-dimensional transform of length $n = 2^m$, we separate even and odd indices:
\begin{equation}
\hat{f}_k = \sum_{j=0}^{n-1} f_j \omega_n^{jk} = \sum_{j=0}^{n/2-1} f_{2j} \omega_n^{2jk} + \sum_{j=0}^{n/2-1} f_{2j+1} \omega_n^{(2j+1)k},
\label{eq:fft_decomposition}
\end{equation}
where $\omega_n = \exp(-2\pi i/n)$ is the primitive $n$-th root of unity. Recognizing that $\omega_n^{2jk} = \omega_{n/2}^{jk}$, we obtain:
\begin{equation}
\hat{f}_k = \sum_{j=0}^{n/2-1} f_{2j} \omega_{n/2}^{jk} + \omega_n^k \sum_{j=0}^{n/2-1} f_{2j+1} \omega_{n/2}^{jk}.
\label{eq:fft_recursive}
\end{equation}

This expresses the $n$-point DFT in terms of two $(n/2)$-point DFTs, enabling recursive application. The twiddle factor $\omega_n^k$ implements the phase shift between even and odd components. For $k \geq n/2$, we use the periodicity $\omega_n^{k+n/2} = -\omega_n^k$ to avoid redundant calculations. The two-dimensional FFT leverages the separability of the exponential kernel:
\begin{equation}
\exp\left(-i\frac{2\pi}{n}(k_x j + k_y k)\right) = \exp\left(-i\frac{2\pi k_x j}{n}\right) \exp\left(-i\frac{2\pi k_y k}{n}\right).
\label{eq:fft_2d_separability}
\end{equation}
This allows sequential application of one-dimensional transforms along each dimension, reducing the 2D problem to $2n$ 1D transforms \cite{Canuto2006}.

The spectral representation of differential operators provides the key advantage of Fourier methods. Taking the Fourier transform of the Laplacian operator in \eqref{eq:final_u_complete}:
\begin{equation}
\mathcal{F}[\nabla^2 u] = \mathcal{F}\left[\frac{\partial^2 u}{\partial x^2} + \frac{\partial^2 u}{\partial y^2}\right].
\label{eq:laplacian_fourier_start}
\end{equation}
Using the differentiation property of Fourier transforms, where $\mathcal{F}[\partial u/\partial x] = ik_x \hat{u}$ \cite{Trefethen2000}:
\begin{equation}
\mathcal{F}[\nabla^2 u] = (ik_x)^2 \hat{u} + (ik_y)^2 \hat{u} = -(k_x^2 + k_y^2)\hat{u} = -k^2 \hat{u}.
\label{eq:laplacian_fourier}
\end{equation}
This remarkable result transforms the differential operator into simple multiplication, eliminating discretization errors associated with finite differences. The spectral accuracy---exponential convergence for smooth functions---arises from this exact representation of derivatives \cite{Tadmor1986}.

The treatment of nonlinear terms requires special consideration. Direct evaluation in spectral space of the cubic term $u^3$ would require a triple convolution:
\begin{equation}
\widehat{u^3}_{k} = \sum_{k_1+k_2+k_3=k} \hat{u}_{k_1} \hat{u}_{k_2} \hat{u}_{k_3}.
\label{eq:convolution_cubic}
\end{equation}
This $O(n^6)$ operation becomes prohibitive for large $n$. The pseudo-spectral approach circumvents this by evaluating nonlinearities in physical space where they reduce to pointwise operations \cite{Orszag1971}. The complete algorithm for advancing the solution from time $t$ to $t + \Delta t$ proceeds as follows:
\begin{enumerate}
\item Transform spectral coefficients to physical space: $u_{jk} = \mathcal{F}^{-1}[\hat{u}_{k_x,k_y}]$
\item Evaluate nonlinear terms pointwise: $f_{jk} = u_{jk} - u_{jk}^3 - u_{jk}v_{jk}^2 + \beta(u_{jk}^2v_{jk} + v_{jk}^3)$
\item Transform nonlinear terms to spectral space: $\hat{f}_{k_x,k_y} = \mathcal{F}[f_{jk}]$
\item Combine diffusion and reaction in spectral space: $d\hat{u}/dt = -D_1 k^2 \hat{u} + \hat{f}$
\end{enumerate}

The implementation as \texttt{rd-spiral} v0.1.0 realizes this mathematical framework through careful software engineering that prioritizes clarity and correctness. The core computational kernel resides in the \texttt{reaction\_diffusion\_rhs} function, which implements the semi-discrete system of ordinary differential equations resulting from spatial discretization. The function signature accepts the current state vector $\mathbf{y} \in \mathbb{C}^{2n^2}$ containing all Fourier coefficients concatenated as $[\hat{u}_{0,0}, \ldots, \hat{u}_{n-1,n-1}, \hat{v}_{0,0}, \ldots, \hat{v}_{n-1,n-1}]^T$, along with the precomputed Laplacian operator eigenvalues $K^2$ and system parameters $(D_1, D_2, \beta)$.

The NumPy implementation leverages the \texttt{numpy.fft} module, which provides optimized C implementations of the Cooley-Tukey algorithm with additional optimizations for non-power-of-two sizes through the Bluestein chirp-z algorithm \cite{Harris2020}. The \texttt{fft2} and \texttt{ifft2} functions automatically handle the proper normalization factors and complex conjugate symmetry required for real-valued fields. Memory layout considerations favor C-contiguous arrays for cache efficiency during the row-wise FFT operations.

The state vector flattening and reshaping operations deserve careful attention to maintain consistency with the mathematical formulation. The \texttt{reshape} operation from the flattened state vector to the 2D spatial grid must preserve the row-major ordering: element $(i,j)$ in the 2D array corresponds to index $i \times n + j$ in the flattened representation. This mapping ensures that the Fourier coefficients maintain their proper spatial interpretation throughout the computation.

Wavenumber array construction follows the NumPy FFT convention, which differs from the mathematical convention by a factor of $2\pi/L$. The physical wavenumbers are computed as:
\begin{lstlisting}[language=Python, basicstyle=\small\ttfamily]
kx = np.fft.fftfreq(n, d=L/n) * 2 * np.pi
ky = np.fft.fftfreq(n, d=L/n) * 2 * np.pi
\end{lstlisting}
The \texttt{fftfreq} function automatically generates the correct ordering $[0, 1, \ldots, n/2-1, -n/2, \ldots, -1]$ scaled by the sampling frequency $n/L$. The subsequent multiplication by $2\pi$ converts to angular wavenumbers. The 2D wavenumber magnitude $K^2 = k_x^2 + k_y^2$ is precomputed using NumPy's broadcasting:
\begin{lstlisting}[language=Python, basicstyle=\small\ttfamily]
kx_grid, ky_grid = np.meshgrid(kx, ky, indexing='ij')
K_squared = kx_grid**2 + ky_grid**2
\end{lstlisting}
The \texttt{indexing='ij'} parameter ensures matrix-style indexing consistent with the mathematical notation where the first index corresponds to $x$ and the second to $y$.

Time integration presents unique challenges for reaction-diffusion systems due to the stiffness introduced by the diffusion operator. The eigenvalues of the discrete Laplacian range from $0$ to $-4\pi^2 n^2/L^2$, creating a condition number that scales as $O(n^2)$. Explicit methods would require timesteps satisfying $\Delta t < O(L^2/(D n^2))$, becoming prohibitively small for fine spatial resolution. While implicit methods offer unconditional stability, they require solving large nonlinear systems at each timestep \cite{Hairer1993}.

Our implementation employs the Dormand-Prince adaptive Runge-Kutta method through SciPy's \texttt{solve\_ivp} interface \cite{Virtanen2020}. This embedded RK5(4) pair provides automatic step size control based on local error estimates \cite{Dormand1980}. The method constructs two approximations of different orders:
\begin{align}
\mathbf{y}_{n+1}^{(4)} &= \mathbf{y}_n + \Delta t \sum_{i=1}^{7} b_i^{(4)} \mathbf{k}_i, \label{eq:rk4_formula}\\
\mathbf{y}_{n+1}^{(5)} &= \mathbf{y}_n + \Delta t \sum_{i=1}^{7} b_i^{(5)} \mathbf{k}_i, \label{eq:rk5_formula}
\end{align}
where the $\mathbf{k}_i$ are intermediate stage evaluations and the $b_i$ coefficients are optimized to minimize error. The local error estimate $\mathbf{e} = \mathbf{y}_{n+1}^{(5)} - \mathbf{y}_{n+1}^{(4)} = O(\Delta t^5)$ guides adaptive step size selection. The controller adjusts the timestep according to:
\begin{equation}
\Delta t_{\text{new}} = \Delta t \cdot \min\left(2, \max\left(0.1, 0.9\left(\frac{\epsilon}{\|\mathbf{e}\|}\right)^{1/5}\right)\right),
\label{eq:step_size_control}
\end{equation}
where $\epsilon$ is the desired tolerance and the safety factors prevent excessive step size variations \cite{Press2007}.

The \texttt{solve\_ivp} function encapsulates this complexity behind a simple interface while providing extensive control over the integration process. Key parameters include \texttt{rtol} (relative tolerance) and \texttt{atol} (absolute tolerance), which control the error per step: $\text{error} \leq \text{atol} + \text{rtol} \times |y|$. For reaction-diffusion systems, we set \texttt{rtol=1e-6} and \texttt{atol=1e-9} to maintain accuracy over long integration times while avoiding excessive computational cost.

Initial condition generation implements the following smooth vortex profile \cite{Bramburger2024}:
\begin{align}
u(x, y, 0) &= \tanh(r) \cos(m\theta - r) \label{eq:initial_u} \\
v(x, y, 0) &= \tanh(r) \sin(m\theta - r) \label{eq:initial_v}
\end{align}
where $r = \sqrt{x^2 + y^2}$ is the radial distance from the origin, $\theta = \arctan2(y, x)$ is the azimuthal angle, and $m$ is the topological charge of the vortex.

These initial conditions \eqref{eq:initial_u}--\eqref{eq:initial_v} are implemented through vectorized NumPy operations that avoid explicit loops. The polar coordinate transformation uses \texttt{numpy.arctan2} rather than \texttt{arctan} to correctly handle all quadrants:
\begin{lstlisting}[language=Python, basicstyle=\small\ttfamily]
r = np.sqrt(X**2 + Y**2)
theta = np.arctan2(Y, X)
u_init = np.tanh(r) * np.cos(m * theta - r)
v_init = np.tanh(r) * np.sin(m * theta - r)
\end{lstlisting}
The hyperbolic tangent provides smooth decay from unity at the origin to zero at infinity, ensuring spectral accuracy. The phase winding $m\theta$ creates the topological charge, while the radial phase shift $-r$ induces rotation \cite{Barkley1991}.

The enhanced solver architecture introduces several improvements over basic implementations. Initial conditions are preserved in compressed NumPy format (\texttt{.npz}) for reproducibility and reference:
\begin{lstlisting}[language=Python, basicstyle=\small\ttfamily]
np.savez_compressed('initial_conditions.npz', 
                   u0=u0, v0=v0, x=grid.x, y=grid.y)
\end{lstlisting}
This binary format maintains full numerical precision while achieving typical compression ratios of 5:1 for smooth fields.

For long-running simulations, the checkpoint mechanism provides resilience against interruptions and enables restart capabilities. When enabled through \texttt{save\_checkpoints=True}, the solver automatically saves intermediate states at specified intervals:
\begin{lstlisting}[language=Python, basicstyle=\small\ttfamily]
checkpoint_times = np.arange(t_start + checkpoint_interval, 
                           t_end, checkpoint_interval)
\end{lstlisting}
Each checkpoint preserves the complete system state in compressed format, allowing simulations to resume from any saved point. This feature proves particularly valuable for the turbulent dynamics configuration, where integration to $t=500$ may require several hours of computation.

Three carefully chosen parameter configurations demonstrate the diversity of dynamics accessible within the model framework. The stable spiral configuration (\texttt{stable\_spiral.txt}) uses symmetric diffusion $D_1 = D_2 = 0.1$ and coupling $\beta = 1.0$, parameters that lie within the stability region for rotating spiral waves. Linear stability analysis of the homogeneous state $(u,v) = (0,0)$ reveals that perturbations with wavenumber $k$ grow at rate $\lambda(k) = 1 - D k^2$, predicting instability for $k < 1/\sqrt{D} \approx 3.16$. The domain size $L = 20$ accommodates several unstable wavelengths, allowing pattern formation while avoiding finite-size effects.

Statistical monitoring throughout the simulation tracks key dynamical indicators. The spatial mean $\langle u \rangle = n^{-2} \sum_{i,j} u_{i,j}$ should remain near zero for the symmetric initial conditions and reaction terms. The standard deviation $\sigma_u = \sqrt{\langle (u - \langle u \rangle)^2 \rangle}$ measures pattern amplitude, converging to a constant value for stable spirals. We compute these statistics at each output timestep using pandas DataFrames for efficient columnar storage and subsequent analysis \cite{McKinney2010}.

The turbulent regime configuration (\texttt{turbulent\_spiral.txt}) employs strongly asymmetric diffusion $D_1 = 0.03$, $D_2 = 0.20$ with reduced coupling $\beta = 0.65$. The extreme diffusion ratio $D_2/D_1 \approx 6.67$ breaks the activator-inhibitor symmetry, promoting spiral breakup and spatiotemporal chaos. The extended domain $L = 50$ on a $256 \times 256$ grid provides sufficient resolution to capture the fine-scale structures and multiple interacting spiral cores that characterize turbulent dynamics. Multiple spiral arms ($m = 4$) in the initial condition enhance the complexity of subsequent evolution. Extended integration to $t = 500$ allows full development of chaotic dynamics, characterized by persistent fluctuations in all statistical measures \cite{Bramburger2024}.

Pattern decay under strong diffusion (\texttt{pattern\_decay.txt}) with $D_1 = D_2 = 0.5$ demonstrates how excessive diffusive transport suppresses pattern formation entirely. Despite favorable reaction parameters ($\beta = 1.0$), the large diffusion coefficients stabilize the homogeneous state against all perturbations. The instability criterion $k < 1/\sqrt{D} \approx 1.41$ combined with the minimum resolvable wavenumber $k_{\min} = 2\pi/L \approx 0.628$ for $L = 10$ predicts marginal instability. Numerical simulations confirm exponential decay to homogeneity with rate consistent with the least stable eigenmode.

The equilibrium analysis feature provides automated characterization of long-time dynamics. By analyzing the final 10\% of the simulation time series (with a minimum of 10 time steps), the solver categorizes the system state based on the temporal behavior of the spatial pattern intensity. For each time step $t$, the spatial standard deviation $\sigma_u(t)$ is computed as:
\begin{equation}
\sigma_u(t) = \sqrt{\frac{1}{n^2} \sum_{i,j=1}^{n} (u_{i,j}(t) - \bar{u}(t))^2}
\end{equation}
where $\bar{u}(t)$ is the spatial mean of the $u$ field at time $t$. The system state is then classified based on the temporal statistics of $\sigma_u$ over the analysis window:
\begin{equation}
\text{State} = \begin{cases}
\text{Homogeneous (pattern decayed)} & \text{if } \langle\sigma_u\rangle < 0.01 \\
\text{Static equilibrium} & \text{if } \text{std}(\sigma_u) < 0.0001 \\
\text{Dynamic equilibrium (steady rotation)} & \text{if } \text{std}(\sigma_u) < 0.001 \\
\text{Quasi-periodic} & \text{if } \text{std}(\sigma_u) < 0.01 \\
\text{Non-equilibrium (chaotic/turbulent)} & \text{otherwise}
\end{cases}
\label{eq:equilibrium_categories}
\end{equation}
where $\langle\sigma_u\rangle$ denotes the temporal mean and $\text{std}(\sigma_u)$ the temporal standard deviation of the spatial pattern intensity over the analysis window. This automated classification enables users to quickly identify the qualitative long-time behavior without manual inspection of time series or spatial patterns.

Output management leverages the HDF5-based NetCDF4 format through xarray, providing self-describing datasets with metadata preservation \cite{Hoyer2017, Rew1990}. The hierarchical structure organizes results as:
\begin{lstlisting}[basicstyle=\small\ttfamily]
solution.nc
+-- dimensions: x, y, time
+-- variables:
|   +-- u(time, x, y): activator concentration
|   +-- v(time, x, y): inhibitor concentration
|   +-- x(x): x-coordinate
|   +-- y(y): y-coordinate
|   +-- time(time): simulation time
+-- attributes:
    +-- d1, d2, beta: model parameters
    +-- n, L: grid parameters
    +-- created: timestamp
\end{lstlisting}
Compression with zlib level 4 typically achieves 10:1 ratios for smooth fields while maintaining lossless precision. The self-documenting nature of NetCDF facilitates data sharing and long-term archival.

The robust error handling ensures that partial results are preserved even if simulations are interrupted or encounter numerical difficulties. A status file documents the simulation outcome:
\begin{lstlisting}[language=Python, basicstyle=\small\ttfamily]
with open('simulation_status.txt', 'w') as f:
    f.write(f"Simulation Status: {status}\n")
    f.write(f"Final time reached: {solution['t'][-1]:.2f}\n")
    f.write(f"Completion: {100*solution['t'][-1]/t_end:.1f}%\n")
\end{lstlisting}
This approach ensures that researchers can assess simulation success and recover useful data from incomplete runs.

Performance profiling using \texttt{cProfile} reveals that FFT operations consume approximately 75\% of computation time, with the remaining 25\% split between memory operations (reshape, concatenate) and the ODE solver overhead. For the $256 \times 256$ turbulent simulation over 5000 timesteps (to $t=500$), typical execution requires approximately 45 minutes on modern hardware (Intel i7-9700K, 16GB RAM), corresponding to $1.5 \times 10^7$ grid point updates per second. This performance, while modest compared to optimized compiled codes, remains acceptable for educational purposes where code clarity takes precedence.

The modular design facilitates experimentation and extension. Users can modify reaction terms by editing the \texttt{reaction\_diffusion\_rhs} function without altering the numerical infrastructure. Alternative initial conditions, boundary conditions (through padding), or even different equations entirely can be accommodated within the same framework. This flexibility, combined with comprehensive inline documentation explaining the correspondence between code and mathematics, creates an effective pedagogical tool for learning numerical methods through hands-on experimentation.

Error analysis confirms the spectral accuracy of the spatial discretization. For smooth solutions, the truncation error decreases exponentially with grid resolution: $\|u - u_n\| \sim \exp(-\alpha n)$ where $\alpha$ depends on the solution regularity. Time integration errors, controlled by the adaptive solver, maintain the specified tolerances throughout the simulation. The combined spatial-temporal accuracy ensures reliable results for the parameter regimes studied, with numerical artifacts remaining below the threshold of physical significance.

\subsection{Data Analysis}

The statistical analysis focused on quantifying the dynamical differences between stable and turbulent spiral wave regimes. For computational efficiency and pedagogical clarity, we restricted our comparison to the stable spiral ($D_1 = D_2 = 0.1$, $\beta = 1.0$) and turbulent ($D_1 = 0.03$, $D_2 = 0.20$, $\beta = 0.65$) configurations, excluding the pattern decay case which exhibits trivial dynamics. All statistical computations were performed using Python 3.11 with NumPy 1.24.3 \cite{Harris2020}, SciPy 1.10.1 \cite{Virtanen2020}, and pandas 2.0.3 \cite{McKinney2010}.

\subsubsection{Time Series Construction}

From the spatiotemporal solution fields $u(\mathbf{x},t)$ and $v(\mathbf{x},t)$, we extracted time series of spatial standard deviations as global measures of pattern intensity:
\begin{align}
\sigma_u(t) &= \sqrt{\frac{1}{n^2} \sum_{i,j=1}^{n} \left[u_{i,j}(t) - \bar{u}(t)\right]^2}, \label{eq:sigma_u_def}\\
\sigma_v(t) &= \sqrt{\frac{1}{n^2} \sum_{i,j=1}^{n} \left[v_{i,j}(t) - \bar{v}(t)\right]^2}, \label{eq:sigma_v_def}
\end{align}
where $\bar{u}(t) = n^{-2}\sum_{i,j} u_{i,j}(t)$ denotes the spatial mean. These scalar time series $\{\sigma_u(t_k), \sigma_v(t_k)\}_{k=1}^{N_t}$ served as the primary observables for statistical analysis, where $N_t$ represents the number of temporal samples until equilibrium was reached in at least one configuration.

\subsubsection{Descriptive Statistics}

For each time series, we computed standard descriptive statistics. The sample mean and unbiased sample variance were calculated as:
\begin{align}
\mu_{\sigma} &= \frac{1}{N_t} \sum_{k=1}^{N_t} \sigma(t_k), \label{eq:sample_mean}\\
s_{\sigma}^2 &= \frac{1}{N_t-1} \sum_{k=1}^{N_t} [\sigma(t_k) - \mu_{\sigma}]^2, \label{eq:sample_var}
\end{align}
where the factor $(N_t-1)^{-1}$ provides an unbiased estimator of the population variance. The coefficient of variation $\text{CV} = s_{\sigma}/\mu_{\sigma}$ provided a dimensionless measure of relative variability. 

Higher-order moments were assessed through the sample skewness and excess kurtosis:
\begin{align}
\gamma_1 &= \frac{1}{N_t} \sum_{k=1}^{N_t} \left(\frac{\sigma(t_k) - \mu_{\sigma}}{s_{\sigma}}\right)^3, \label{eq:skewness}\\
\gamma_2 &= \frac{1}{N_t} \sum_{k=1}^{N_t} \left(\frac{\sigma(t_k) - \mu_{\sigma}}{s_{\sigma}}\right)^4 - 3. \label{eq:excess_kurtosis}
\end{align}
The robust location and scale measures included the median, interquartile range (IQR = $Q_3 - Q_1$), and median absolute deviation:
\begin{equation}
\text{MAD} = \text{median}(|\sigma(t_k) - \text{median}(\sigma)|). \label{eq:mad}
\end{equation}

\subsubsection{Normality Assessment}

Given the nonlinear dynamics inherent in reaction-diffusion systems, we conducted rigorous normality testing using multiple complementary approaches. Each test examines different aspects of the distribution, providing a comprehensive assessment of departures from Gaussianity.

\paragraph{Shapiro-Wilk Test} The Shapiro-Wilk test \cite{Shapiro1965} assesses normality by examining the linearity of the quantile-quantile plot. Let $X_{(1)} \leq X_{(2)} \leq \cdots \leq X_{(N_t)}$ denote the order statistics of our sample. The test statistic is constructed as:
\begin{equation}
W = \frac{\left(\sum_{i=1}^{N_t} a_i X_{(i)}\right)^2}{\sum_{i=1}^{N_t} (X_i - \bar{X})^2}, \label{eq:shapiro_wilk_def}
\end{equation}
where the coefficients $a_i$ are obtained from the expected values of standard normal order statistics.

To derive these coefficients, consider the covariance matrix $\mathbf{V}$ of the order statistics from a standard normal distribution, with elements:
\begin{equation}
V_{ij} = \text{Cov}(Z_{(i)}, Z_{(j)}) = \mathbb{E}[Z_{(i)}Z_{(j)}] - \mathbb{E}[Z_{(i)}]\mathbb{E}[Z_{(j)}], \label{eq:order_stat_cov}
\end{equation}
where $Z_{(i)}$ denotes the $i$-th order statistic from a standard normal sample of size $N_t$.

Let $\mathbf{m} = (\mathbb{E}[Z_{(1)}], \mathbb{E}[Z_{(2)}], \ldots, \mathbb{E}[Z_{(N_t)}])^T$ be the vector of expected values of standard normal order statistics. The optimal coefficients are given by:
\begin{equation}
\mathbf{a} = \frac{\mathbf{m}^T\mathbf{V}^{-1}}{\sqrt{\mathbf{m}^T\mathbf{V}^{-1}\mathbf{m}}}. \label{eq:shapiro_coefficients}
\end{equation}

The numerator in \eqref{eq:shapiro_wilk_def} represents the best linear unbiased estimator of the scale parameter under normality, while the denominator is the usual sum of squares. Under the null hypothesis of normality, $W$ approaches 1, with smaller values indicating departures from normality.

\paragraph{D'Agostino-Pearson Omnibus Test} The D'Agostino-Pearson test \cite{DAgostino1990} combines separate tests for skewness and kurtosis into an omnibus test statistic. Starting with the sample skewness from \eqref{eq:skewness}, we first transform it to achieve approximate normality.

Define the intermediate quantities:
\begin{align}
\beta_2(\gamma_1) &= \frac{3(N_t^2 + 27N_t - 70)(N_t + 1)(N_t + 3)}{(N_t - 2)(N_t + 5)(N_t + 7)(N_t + 9)}, \label{eq:beta2_g1}\\
W^2 &= -1 + \sqrt{2(\beta_2(\gamma_1) - 1)}, \label{eq:W_squared}\\
\delta &= 1/\sqrt{\ln W}, \label{eq:delta_transform}\\
\alpha &= \sqrt{\frac{2}{W^2 - 1}}. \label{eq:alpha_transform}
\end{align}

The standardized skewness statistic is then:
\begin{equation}
Z_{\gamma_1} = \delta \ln\left(\frac{\gamma_1}{\alpha} + \sqrt{\left(\frac{\gamma_1}{\alpha}\right)^2 + 1}\right), \label{eq:z_skewness}
\end{equation}
which approximately follows a standard normal distribution under the null hypothesis.

For kurtosis, we similarly transform the sample excess kurtosis $\gamma_2$ from \eqref{eq:excess_kurtosis}. First, compute the expected value and variance of $\gamma_2$ under normality:
\begin{align}
\mathbb{E}[\gamma_2] &= \frac{-6}{N_t + 1}, \label{eq:expected_kurtosis}\\
\text{Var}(\gamma_2) &= \frac{24N_t(N_t - 2)(N_t - 3)}{(N_t + 1)^2(N_t + 3)(N_t + 5)}. \label{eq:var_kurtosis}
\end{align}

The standardized kurtosis is:
\begin{equation}
x = \frac{\gamma_2 - \mathbb{E}[\gamma_2]}{\sqrt{\text{Var}(\gamma_2)}}. \label{eq:standardized_kurtosis}
\end{equation}

To achieve better normality, we apply a further transformation. Define:
\begin{align}
\beta_2(\gamma_2) &= \frac{6(N_t^2 - 5N_t + 2)}{(N_t + 7)(N_t + 9)} \sqrt{\frac{6(N_t + 3)(N_t + 5)}{N_t(N_t - 2)(N_t - 3)}}, \label{eq:beta2_g2}\\
A &= 6 + \frac{8}{\beta_2(\gamma_2)}\left(\frac{2}{\beta_2(\gamma_2)} + \sqrt{1 + \frac{4}{\beta_2(\gamma_2)^2}}\right). \label{eq:A_parameter}
\end{align}

The transformed kurtosis statistic is:
\begin{equation}
Z_{\gamma_2} = \sqrt{\frac{9A}{2}}\left[1 - \frac{2}{9A} - \left(\frac{1 - 2/A}{1 + x\sqrt{2/(A-4)}}\right)^{1/3}\right], \label{eq:z_kurtosis}
\end{equation}
which also approximately follows a standard normal distribution under normality.

The omnibus test statistic combines both components:
\begin{equation}
K^2 = Z_{\gamma_1}^2 + Z_{\gamma_2}^2, \label{eq:omnibus_statistic}
\end{equation}
which follows a $\chi^2_2$ distribution under the null hypothesis of normality.

\paragraph{Anderson-Darling Test} The Anderson-Darling test \cite{Anderson1952} measures the distance between the empirical distribution function and the hypothesized normal distribution, with greater weight given to the tails. First, standardize the observations:
\begin{equation}
Z_i = \frac{X_i - \bar{X}}{s}, \quad i = 1, 2, \ldots, N_t, \label{eq:standardize_obs}
\end{equation}
where $\bar{X}$ and $s$ are the sample mean and standard deviation.

Order the standardized values: $Z_{(1)} \leq Z_{(2)} \leq \cdots \leq Z_{(N_t)}$. The test statistic is:
\begin{equation}
A^2 = -N_t - \frac{1}{N_t}\sum_{i=1}^{N_t} (2i - 1)[\ln \Phi(Z_{(i)}) + \ln(1 - \Phi(Z_{(N_t+1-i)}))], \label{eq:anderson_darling_stat}
\end{equation}
where $\Phi(\cdot)$ is the standard normal cumulative distribution function.

To understand this construction, note that under the null hypothesis, $\Phi(Z_{(i)})$ should be approximately uniformly distributed. The statistic can be rewritten as:
\begin{equation}
A^2 = -N_t - \frac{1}{N_t}\sum_{i=1}^{N_t} (2i - 1)\ln[U_{(i)}(1 - U_{(N_t+1-i)})], \label{eq:ad_uniform}
\end{equation}
where $U_{(i)} = \Phi(Z_{(i)})$. The weights $(2i - 1)$ give more importance to deviations in the tails of the distribution.

The null distribution of $A^2$ depends on the sample size and the fact that parameters are estimated from the data. Critical values are obtained through extensive simulation \cite{Stephens1974}.

\paragraph{Jarque-Bera Test} The Jarque-Bera test \cite{Jarque1987} is based on the fact that a normal distribution has skewness 0 and excess kurtosis 0. Using the sample moments from \eqref{eq:skewness} and \eqref{eq:excess_kurtosis}, the test statistic is:
\begin{equation}
\text{JB} = \frac{N_t}{6}\left(\gamma_1^2 + \frac{\gamma_2^2}{4}\right). \label{eq:jarque_bera_stat}
\end{equation}

To derive the null distribution, note that under normality and for large $N_t$:
\begin{align}
\sqrt{N_t}\gamma_1 &\stackrel{d}{\rightarrow} \mathcal{N}\left(0, 6\right), \label{eq:asymp_skewness}\\
\sqrt{N_t}\gamma_2 &\stackrel{d}{\rightarrow} \mathcal{N}\left(0, 24\right), \label{eq:asymp_kurtosis}
\end{align}
where $\stackrel{d}{\rightarrow}$ denotes convergence in distribution.

Therefore:
\begin{align}
\frac{N_t\gamma_1^2}{6} &\stackrel{d}{\rightarrow} \chi^2_1, \label{eq:chi2_skewness}\\
\frac{N_t\gamma_2^2}{24} &\stackrel{d}{\rightarrow} \chi^2_1. \label{eq:chi2_kurtosis}
\end{align}

Since $\gamma_1$ and $\gamma_2$ are asymptotically independent under normality, the test statistic:
\begin{equation}
\text{JB} = \frac{N_t\gamma_1^2}{6} + \frac{N_t\gamma_2^2}{24} \stackrel{d}{\rightarrow} \chi^2_2. \label{eq:jb_chi2}
\end{equation}

Thus, under the null hypothesis of normality, the Jarque-Bera statistic asymptotically follows a chi-squared distribution with 2 degrees of freedom.

\subsubsection{Bootstrap Methods}

The bootstrap method \cite{Efron1979} provides a powerful framework for statistical inference without parametric assumptions. We present a rigorous mathematical foundation for the bootstrap procedure employed in our analysis.

\paragraph{Bootstrap Principle} Let $\mathbf{X} = (X_1, X_2, \ldots, X_{N_t})$ be our observed sample from an unknown distribution $F$. We wish to estimate the sampling distribution of a statistic $\theta = T(\mathbf{X}, F)$. The empirical distribution function:
\begin{equation}
\hat{F}_n(x) = \frac{1}{N_t} \sum_{i=1}^{N_t} \mathbf{1}(X_i \leq x), \label{eq:empirical_cdf}
\end{equation}
serves as our estimate of $F$, where $\mathbf{1}(\cdot)$ is the indicator function.

The bootstrap principle replaces the unknown $F$ with $\hat{F}_n$ and approximates the sampling distribution of $\theta$ by the distribution of $\theta^* = T(\mathbf{X}^*, \hat{F}_n)$, where $\mathbf{X}^*$ is a sample drawn from $\hat{F}_n$.

\paragraph{Bootstrap Algorithm} Drawing from $\hat{F}_n$ is equivalent to sampling with replacement from the original data. Define the bootstrap sample:
\begin{equation}
\mathbf{X}^{*b} = (X_{I_1}^*, X_{I_2}^*, \ldots, X_{I_{N_t}}^*), \label{eq:bootstrap_sample}
\end{equation}
where $I_1, I_2, \ldots, I_{N_t}$ are independent random variables uniformly distributed on $\{1, 2, \ldots, N_t\}$.

For each bootstrap iteration $b = 1, 2, \ldots, B$ with $B = 5{,}000$:
\begin{enumerate}
\item Generate indices $I_1^{(b)}, \ldots, I_{N_t}^{(b)} \stackrel{\text{iid}}{\sim} \text{Uniform}\{1, \ldots, N_t\}$
\item Construct $\mathbf{X}^{*b} = (X_{I_1^{(b)}}, \ldots, X_{I_{N_t}^{(b)}})$
\item Compute $\theta^{*b} = T(\mathbf{X}^{*b})$
\end{enumerate}

The bootstrap distribution is the empirical distribution of $\{\theta^{*1}, \theta^{*2}, \ldots, \theta^{*5000}\}$.

\paragraph{Bootstrap Confidence Intervals} For a confidence level $1-\alpha$, the percentile method constructs the interval:
\begin{equation}
\text{CI}_{\text{percentile}} = [\theta_{\alpha/2}^*, \theta_{1-\alpha/2}^*], \label{eq:percentile_ci}
\end{equation}
where $\theta_{\beta}^*$ is the $\beta$-quantile of the bootstrap distribution:
\begin{equation}
\theta_{\beta}^* = \inf\{t : \frac{1}{5000}\sum_{b=1}^{5000} \mathbf{1}(\theta^{*b} \leq t) \geq \beta\}. \label{eq:bootstrap_quantile}
\end{equation}

\paragraph{Bootstrap for Normality Tests} For each normality test statistic $T_{\text{norm}}$ (e.g., Shapiro-Wilk $W$), we assessed stability through:
\begin{equation}
\hat{p}_{\text{normal}} = \frac{1}{5000} \sum_{b=1}^{5000} \mathbf{1}(p^{*b} > 0.05), \label{eq:bootstrap_normality_prop}
\end{equation}
where $p^{*b}$ is the $p$-value obtained from applying the test to bootstrap sample $b$. This proportion estimates the probability that a resampled dataset would suggest normality at the 5\% significance level.

\subsubsection{Non-parametric Hypothesis Testing}

The consistent rejection of normality necessitated distribution-free methods for comparing stable and turbulent dynamics.

\paragraph{Mann-Whitney U Test} The Mann-Whitney U test \cite{Mann1947} tests the null hypothesis $H_0: F_X = F_Y$ against the alternative $H_1: F_X(t) = F_Y(t - \Delta)$ for some shift $\Delta \neq 0$.

Given samples $\mathbf{X} = (X_1, \ldots, X_{n_1})$ and $\mathbf{Y} = (Y_1, \ldots, Y_{n_2})$, combine them into a single ordered sample and assign ranks $R_1, \ldots, R_{n_1+n_2}$. Let $R_{X,i}$ denote the rank of $X_i$ in the combined sample. The Mann-Whitney statistic is:
\begin{equation}
U_X = \sum_{i=1}^{n_1} R_{X,i} - \frac{n_1(n_1 + 1)}{2}. \label{eq:mann_whitney_u}
\end{equation}

Equivalently, $U_X$ counts the number of pairs $(X_i, Y_j)$ where $X_i > Y_j$:
\begin{equation}
U_X = \sum_{i=1}^{n_1} \sum_{j=1}^{n_2} \mathbf{1}(X_i > Y_j). \label{eq:mann_whitney_count}
\end{equation}

Under $H_0$, the expected value and variance are:
\begin{align}
\mathbb{E}[U_X] &= \frac{n_1 n_2}{2}, \label{eq:u_expectation}\\
\text{Var}(U_X) &= \frac{n_1 n_2 (n_1 + n_2 + 1)}{12}. \label{eq:u_variance}
\end{align}

For large samples, the standardized statistic:
\begin{equation}
Z = \frac{U_X - \mathbb{E}[U_X]}{\sqrt{\text{Var}(U_X)}} \stackrel{d}{\rightarrow} \mathcal{N}(0, 1), \label{eq:mann_whitney_z}
\end{equation}
follows an approximate standard normal distribution.

When ties are present, the variance requires adjustment:
\begin{equation}
\text{Var}_{\text{tie}}(U_X) = \frac{n_1 n_2}{12}\left(n_1 + n_2 + 1 - \frac{\sum_{g=1}^{G} t_g(t_g^2 - 1)}{(n_1 + n_2)(n_1 + n_2 - 1)}\right), \label{eq:u_variance_ties}
\end{equation}
where $G$ is the number of tied groups and $t_g$ is the size of the $g$-th tied group.

\paragraph{Kolmogorov-Smirnov Test} The two-sample Kolmogorov-Smirnov test \cite{Smirnov1948} is sensitive to any difference in distributions, not just location shifts. Define the empirical distribution functions:
\begin{align}
F_{1,n_1}(x) &= \frac{1}{n_1} \sum_{i=1}^{n_1} \mathbf{1}(X_i \leq x), \label{eq:ecdf_x}\\
F_{2,n_2}(x) &= \frac{1}{n_2} \sum_{j=1}^{n_2} \mathbf{1}(Y_j \leq x). \label{eq:ecdf_y}
\end{align}

The test statistic is the supremum of the absolute difference:
\begin{equation}
D_{n_1,n_2} = \sup_{x \in \mathbb{R}} |F_{1,n_1}(x) - F_{2,n_2}(x)|. \label{eq:ks_supremum}
\end{equation}

In practice, the supremum is achieved at one of the observed values. Combining and sorting all observations: $Z_{(1)} \leq Z_{(2)} \leq \cdots \leq Z_{(n_1+n_2)}$, we compute:
\begin{equation}
D_{n_1,n_2} = \max_{k=1,\ldots,n_1+n_2} \left|F_{1,n_1}(Z_{(k)}) - F_{2,n_2}(Z_{(k)})\right|. \label{eq:ks_computational}
\end{equation}

Under the null hypothesis of equal distributions, the limiting distribution of $D_{n_1,n_2}$ is given by:
\begin{equation}
\lim_{n_1,n_2 \to \infty} P\left(\sqrt{\frac{n_1 n_2}{n_1 + n_2}} D_{n_1,n_2} \leq t\right) = K(t), \label{eq:ks_limit_dist}
\end{equation}
where $K(t)$ is the Kolmogorov distribution function:
\begin{equation}
K(t) = 1 - 2\sum_{j=1}^{\infty} (-1)^{j-1} e^{-2j^2t^2}. \label{eq:kolmogorov_cdf}
\end{equation}

\subsubsection{Effect Size Quantification}

\paragraph{Cliff's Delta} Cliff's delta \cite{Cliff1993} provides a non-parametric measure of how often values from one group exceed values from the other. For samples $\mathbf{X} = (X_1, \ldots, X_{n_1})$ and $\mathbf{Y} = (Y_1, \ldots, Y_{n_2})$, define the dominance matrix:
\begin{equation}
d_{ij} = \begin{cases}
+1 & \text{if } X_i > Y_j \\
0 & \text{if } X_i = Y_j \\
-1 & \text{if } X_i < Y_j
\end{cases}. \label{eq:dominance_matrix}
\end{equation}

Cliff's delta is the mean of the dominance matrix:
\begin{equation}
\delta = \frac{1}{n_1 n_2} \sum_{i=1}^{n_1} \sum_{j=1}^{n_2} d_{ij}. \label{eq:cliff_delta_mean}
\end{equation}

This can be expressed in terms of probabilities:
\begin{equation}
\delta = P(X > Y) - P(X < Y) = 2P(X > Y) - 1 + P(X = Y). \label{eq:cliff_delta_prob}
\end{equation}

The connection to the Mann-Whitney U statistic is:
\begin{equation}
\delta = \frac{2U_X}{n_1 n_2} - 1, \label{eq:cliff_mann_whitney_relation}
\end{equation}
when there are no ties.

\paragraph{Variance and Confidence Intervals} The variance of Cliff's delta under the null hypothesis involves fourth-order moments. For large samples, we employ the bootstrap for inference. The bootstrap algorithm for Cliff's delta with $B = 5{,}000$ iterations:

\begin{enumerate}
\item For iteration $b = 1, \ldots, 5{,}000$:
   \begin{enumerate}
   \item Draw $\mathbf{X}^{*b} = (X_{I_1}^*, \ldots, X_{I_{n_1}}^*)$ with replacement from $\mathbf{X}$
   \item Draw $\mathbf{Y}^{*b} = (Y_{J_1}^*, \ldots, Y_{J_{n_2}}^*)$ with replacement from $\mathbf{Y}$
   \item Compute $\delta^{*b}$ using \eqref{eq:cliff_delta_mean} on the bootstrap samples
   \end{enumerate}
\item Construct the percentile confidence interval:
   \begin{equation}
   \text{CI}_{95\%}(\delta) = [\delta_{0.025}^*, \delta_{0.975}^*]. \label{eq:cliff_ci}
   \end{equation}
\end{enumerate}

The bootstrap standard error is:
\begin{equation}
\text{SE}_{\text{boot}}(\delta) = \sqrt{\frac{1}{4999} \sum_{b=1}^{5000} (\delta^{*b} - \bar{\delta}^*)^2}, \label{eq:bootstrap_se}
\end{equation}
where $\bar{\delta}^* = 5000^{-1}\sum_{b=1}^{5000} \delta^{*b}$.

\subsubsection{Spatial Entropy Analysis}

To quantify the spatial organization and complexity of spiral wave patterns, we employed a comprehensive information-theoretic framework based on Shannon entropy \cite{Shannon1948}. For reaction-diffusion systems exhibiting pattern formation, entropy measures provide quantitative discrimination between ordered spiral rotation and spatiotemporal chaos \cite{Cross1993,Kuramoto2003}.

For a continuous concentration field $\phi(\mathbf{x},t)$ where $\phi \in \{u, v, u+v\}$, the differential entropy is defined as:
\begin{equation}
h[\phi] = -\int_{\mathbb{R}} f_\phi(\xi) \log f_\phi(\xi) \, d\xi, \label{eq:differential_entropy}
\end{equation}
where $f_\phi(\xi)$ is the probability density function of field values. In practice, we approximate this continuous measure through discretization.

First, we normalize each field to the unit interval to ensure scale invariance:
\begin{equation}
\tilde{\phi}_{i,j}(t) = \frac{\phi_{i,j}(t) - \min_{k,l} \phi_{k,l}(t)}{\max_{k,l} \phi_{k,l}(t) - \min_{k,l} \phi_{k,l}(t)}, \label{eq:normalize_field_entropy}
\end{equation}
where the indices $(i,j)$ span the computational grid.

Following the maximum entropy principle \cite{Jaynes1957}, we discretize the normalized field into $M = 256$ equiprobable bins under the null hypothesis of uniformity. The bin boundaries are defined as $b_k = k/M$ for $k = 0, 1, \ldots, M$, and we compute the empirical histogram:
\begin{equation}
N_k = \sum_{i=1}^{n}\sum_{j=1}^{n} \mathbf{1}(b_{k-1} \leq \tilde{\phi}_{i,j} < b_k), \label{eq:bin_counts}
\end{equation}
where $\mathbf{1}(\cdot)$ is the indicator function. The empirical probability mass function is then $p_k = N_k/n^2$.

The discrete Shannon entropy, measured in bits, becomes:
\begin{equation}
H[\phi(t)] = -\sum_{k=1}^{M} p_k \log_2 p_k, \label{eq:discrete_entropy}
\end{equation}
where we adopt the convention that $0 \log_2 0 = 0$.

To capture spatial heterogeneity beyond global statistics, we computed the local entropy through a sliding window approach. For each spatial location $(i,j)$, we define a local neighborhood $\mathcal{N}_{i,j}$ of size $w \times w$ (with $w = 5$ in our implementation) and calculate:
\begin{equation}
H_{\text{local}}(i,j,t) = -\sum_{k=1}^{K} p_{k}^{(i,j)} \log_2 p_{k}^{(i,j)}, \label{eq:local_entropy}
\end{equation}
where $p_{k}^{(i,j)}$ is the probability distribution within the local window and $K < M$ reflects the reduced number of bins for robust estimation in small samples. The mean spatial entropy is then:
\begin{equation}
\langle H_{\text{spatial}} \rangle(t) = \frac{1}{n^2} \sum_{i,j} H_{\text{local}}(i,j,t). \label{eq:mean_spatial_entropy}
\end{equation}

For the coupled reaction-diffusion system, we further quantified the statistical dependence between the $u$ and $v$ fields through the joint entropy:
\begin{equation}
H(U,V) = -\sum_{i=1}^{M}\sum_{j=1}^{M} p_{ij} \log_2 p_{ij}, \label{eq:joint_entropy}
\end{equation}
where $p_{ij}$ is the joint probability mass function obtained from the two-dimensional histogram of $(u,v)$ values.

The mutual information, measuring the reduction in uncertainty about one field given knowledge of the other, follows as:
\begin{equation}
I(U;V) = H(U) + H(V) - H(U,V), \label{eq:mutual_information}
\end{equation}
which quantifies the coupling strength between the activator and inhibitor fields. The normalized mutual information:
\begin{equation}
I_{\text{norm}}(U;V) = \frac{I(U;V)}{\sqrt{H(U)H(V)}}, \label{eq:normalized_mi}
\end{equation}
provides a scale-invariant measure in $[0,1]$, with higher values indicating stronger coupling.

For interpretation, the maximum entropy $H_{\max} = \log_2 M = 8$ bits corresponds to a uniform distribution (maximum disorder), while $H_{\min} = 0$ indicates complete spatial homogeneity. In the context of spiral waves, stable rotation exhibits intermediate entropy values with low temporal variance, while turbulent dynamics show elevated entropy with persistent fluctuations, providing a quantitative measure of the transition from order to spatiotemporal chaos \cite{Davidenko1992}.

\section{Results}

\subsection{Computational Experiments and System Evolution}

Three distinct parameter regimes were systematically investigated to characterize the full spectrum of reaction-diffusion dynamics accessible within the model framework. Each simulation employed the enhanced pseudo-spectral solver with adaptive time integration, executing on identical computational infrastructure to ensure methodological consistency. The experimental protocol prioritized comprehensive temporal resolution while maintaining numerical stability throughout extended integration periods.

\subsubsection{Experiment 1: Stable Spiral Dynamics}

The stable spiral configuration ($D_1 = D_2 = 0.1$, $\beta = 1.0$) was initialized on a $128 \times 128$ computational grid spanning a physical domain of $L = 20$ spatial units. Integration proceeded from $t = 0$ to $t = 200$ with temporal resolution $\Delta t = 0.1$, yielding 2,001 discrete time points for analysis. The simulation commenced at 22:36:33 UTC and reached completion at 22:40:01 UTC, requiring approximately 3.5 minutes of wall-clock time.

Throughout the integration process, the solver demonstrated exceptional numerical efficiency, with the effective time-stepping rate accelerating from an initial 0.33 time units per second during the transient phase to a sustained rate exceeding 1.13 time units per second once the spiral structure stabilized. This performance acceleration reflected the adaptive integrator's ability to increase step sizes as the solution approached its asymptotic state. The integration achieved 98.6\% completion before the final rapid convergence phase, with no numerical instabilities or convergence failures reported.

Real-time monitoring revealed systematic evolution toward dynamic equilibrium. The spatial standard deviation of the $u$-field converged to $\sigma_u = 0.670790 \pm 0.000155$ over the final 200 time steps, exhibiting minimal variation (coefficient of variation = 0.0232\%). This remarkable stability, characterized by a pattern range of merely 0.000466 concentration units, confirmed the establishment of a steadily rotating spiral wave with constant angular velocity and preserved topological structure.

\subsubsection{Experiment 2: Turbulent Spiral Dynamics}

The turbulent regime investigation employed significantly enhanced computational resources to capture the complex spatiotemporal dynamics emerging from asymmetric diffusion coefficients ($D_1 = 0.03$, $D_2 = 0.20$) and reduced coupling strength ($\beta = 0.65$). The simulation utilized a $256 \times 256$ grid discretizing an extended domain of $L = 50$ spatial units, necessitating 65,536 grid points compared to the 16,384 points in the stable spiral experiment. Integration spanned from $t = 0$ to $t = 500$ with consistent temporal resolution $\Delta t = 0.1$, generating 5,001 temporal samples.

Given the anticipated computational demands of this extended simulation, the checkpoint mechanism was activated with 100 time unit intervals, creating five intermediate state saves throughout the integration. The simulation initiated at 21:21:50 UTC and concluded at 22:42:39 UTC, requiring 1 hour 20 minutes 49 seconds of continuous computation. This represented a 23-fold increase in computational time despite only a 2.5-fold increase in temporal extent, reflecting both the quadrupled spatial resolution and the increased stiffness arising from the disparate diffusion coefficients.

The integration exhibited markedly different performance characteristics compared to the stable spiral case. Initial time-stepping rates of 0.02 time units per second gradually increased to a maximum of 0.16 time units per second, but with substantial fluctuations throughout the simulation. The checkpoint mechanism proved essential, with successful saves at $t = \{100, 200, 300, 400, 500\}$, each requiring approximately 12 seconds for data serialization. Between checkpoints, the average integration rate stabilized near 0.12 time units per second, approximately one-tenth the peak performance observed in the stable spiral simulation.

Notably, equilibrium analysis was deliberately disabled for this configuration based on \emph{a priori} knowledge of the system's chaotic nature. The simulation status explicitly recorded "COMPLETE - No equilibrium analysis requested", acknowledging that conventional equilibrium metrics would be meaningless for a system exhibiting persistent spatiotemporal chaos.

\subsubsection{Experiment 3: Pattern Decay Dynamics}

The pattern decay experiment investigated the critical regime where excessive diffusion ($D_1 = D_2 = 0.5$) overwhelms the reaction dynamics, preventing sustained pattern formation. This configuration employed a reduced $64 \times 64$ grid on a compact domain of $L = 10$ spatial units, optimized for capturing the relatively simple decay dynamics. Integration proceeded from $t = 0$ to $t = 100$ with standard temporal resolution $\Delta t = 0.1$, generating 1,001 time points.

Initiated at 22:43:20 UTC and completing at 22:46:02 UTC, this simulation required only 2 minutes 42 seconds of computational time, demonstrating the efficiency gains possible when modeling systems with smooth, monotonic evolution. The time-stepping rate exhibited characteristic behavior for a decaying system: initial rates of 0.60--0.65 time units per second during active pattern dissolution, followed by acceleration to sustained rates exceeding 0.65 time units per second as the system approached homogeneity.

The equilibrium analysis, conducted over the final 100 time steps, quantified the extent of pattern decay with exceptional precision. The spatial standard deviation collapsed to $\sigma_u = 0.000006 \pm 0.000005$, representing a five order-of-magnitude reduction from typical pattern amplitudes. The pattern range of 0.000016 concentration units approached the numerical precision limits of the double-precision arithmetic. Despite the near-zero mean intensity, the relative variation remained substantial (76.0\%), reflecting the dominance of numerical noise in the absence of coherent spatial structure.

The automated equilibrium classification algorithm correctly identified the final state as "HOMOGENEOUS (pattern decayed)", confirming complete suppression of spatial organization under the strong diffusion regime. This classification was reached through systematic evaluation of pattern intensity metrics falling below the predetermined threshold of 0.01, triggering the homogeneous state designation in the diagnostic framework.

\subsection{Statistical Analyses}

The temporal evolution of spiral wave dynamics revealed fundamentally different statistical signatures between stable and turbulent regimes. In the stable spiral configuration, both concentration fields exhibited remarkably consistent behavior throughout the simulation. The spatial standard deviation of the $u$-field maintained a mean value of $\sigma_u = 0.670465 \pm 0.001711$, with median 0.670638 and an exceptionally narrow interquartile range of 0.000255. Similarly, the $v$-field showed virtually identical statistical properties ($\sigma_v = 0.670464 \pm 0.001710$, median 0.670637, IQR 0.000263), confirming the symmetric nature of the stable dynamics.

The turbulent regime, by contrast, demonstrated substantially enhanced variability in both fields. The $u$-field standard deviation averaged $0.665332 \pm 0.011852$ with median 0.665146 and interquartile range 0.015637, while the $v$-field showed even greater fluctuations ($\sigma_v = 0.654919 \pm 0.012626$, median 0.654612, IQR 0.016164). These differences translated to dramatic increases in the coefficient of variation: from 0.0026 in the stable regime to 0.0178 for $\sigma_u$ and 0.0193 for $\sigma_v$ in the turbulent case, representing approximately 7-fold amplification of relative variability.

\begin{figure}[H]
    \centering
    \includegraphics[width=\linewidth]{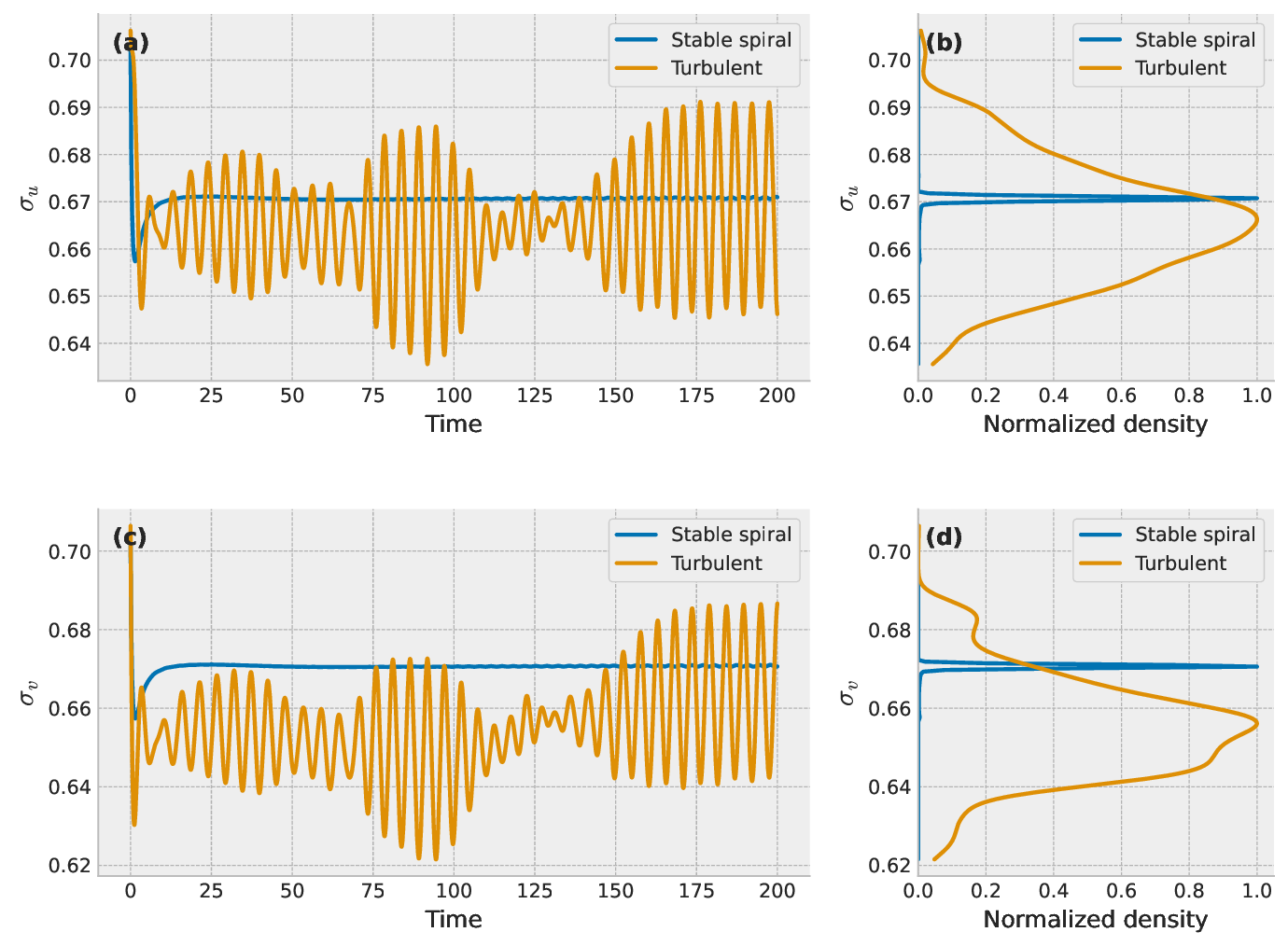}
    \caption{Temporal evolution of spatial standard deviations for stable and turbulent spiral wave dynamics. (a) Time series of $\sigma_u$ showing constant amplitude for stable spiral (blue) and fluctuating amplitude for turbulent spiral (orange). (b) Normalized density distributions of $\sigma_u$ revealing narrow peaked distribution for stable and broad distribution for turbulent dynamics. (c,d) Corresponding plots for $\sigma_v$ exhibiting similar characteristics. Parameters: stable spiral ($D_1 = D_2 = 0.1$, $\beta = 1.0$), turbulent spiral ($D_1 = 0.03$, $D_2 = 0.20$, $\beta = 0.65$).}
    \label{fig:f1}
\end{figure}

Given the complex nonlinear dynamics inherent in reaction-diffusion systems, we conducted comprehensive normality testing using four complementary approaches, each validated through 5,000 bootstrap iterations. The stable spiral series exhibited extreme departures from Gaussian behavior across all metrics. The Shapiro-Wilk test for $\sigma_u$ yielded $W = 0.254894$ ($p < 0.001$), with bootstrap analysis confirming this result (mean $W = 0.256598 \pm 0.020050$, 95\% CI: [0.215915, 0.292351]). The D'Agostino-Pearson omnibus test, which combines skewness and kurtosis assessments, produced an even more decisive rejection with $K^2 = 811.076404$ ($p < 0.001$). Bootstrap validation revealed substantial variability in this statistic (mean $K^2 = 1585.998444 \pm 552.044741$) but consistent non-normality across all resamples.

The Anderson-Darling test, particularly sensitive to deviations in distribution tails, generated $A^2 = 520.790013$, far exceeding the 5\% critical value of 0.785000. Bootstrap analysis confirmed this extreme value was representative (mean $A^2 = 519.872150 \pm 12.743832$). Most strikingly, the Jarque-Bera test produced $JB = 821290.009447$ ($p < 0.001$), with bootstrap estimates showing enormous variability (mean $JB = 765762.512642 \pm 561548.392272$) but unanimous rejection of normality. Remarkably, zero percent of bootstrap samples suggested normality across all four tests, providing robust evidence against Gaussian behavior.

The turbulent spiral series, while still significantly non-normal, showed less extreme departures from Gaussianity. The Shapiro-Wilk test yielded $W = 0.995543$ ($p = 0.000011$), much closer to unity than the stable case, with tight bootstrap confidence intervals (mean $W = 0.994825 \pm 0.001101$, CI: [0.992284, 0.996637]). The D'Agostino-Pearson test produced $K^2 = 13.319725$ ($p = 0.001281$), with 2.6\% of bootstrap samples failing to reject normality. The Anderson-Darling statistic ($A^2 = 1.644380$) still exceeded critical values, though only 0.5\% of bootstrap samples suggested normality. The Jarque-Bera test gave similar results ($JB = 13.512541$, $p = 0.001164$), with 3.0\% of bootstrap samples consistent with normality.

Analysis of distributional shape parameters revealed the mechanistic basis for these normality violations. The stable series exhibited extreme leptokurtosis, with excess kurtosis values of 96.2497 for $\sigma_u$ and 96.4037 for $\sigma_v$, indicating highly peaked distributions with minimal variation around the mean. Conversely, the turbulent series showed platykurtic distributions (excess kurtosis: -3.0984 for $\sigma_u$, -2.7573 for $\sigma_v$), reflecting broader, flatter distributions characteristic of systems exploring a wider range of dynamical states.

Non-parametric hypothesis testing provided robust evidence for significant differences between regimes without distributional assumptions. The Mann-Whitney U test, assessing differences in central tendency, yielded $U = 2749725.0$ ($p < 0.001$) for $\sigma_u$ comparison and $U = 3566728.0$ ($p < 0.001$) for $\sigma_v$. The Kolmogorov-Smirnov test, sensitive to any distributional differences, produced even stronger evidence with $D = 0.629685$ for $\sigma_u$ and $D = 0.841079$ for $\sigma_v$ (both $p < 0.001$).

To quantify effect magnitudes, we computed Cliff's delta with bootstrap confidence intervals. For $\sigma_u$, we obtained $\delta = 0.3735$ (95\% CI: [0.3354, 0.4117]), indicating a medium effect size. The $\sigma_v$ comparison revealed a substantially larger effect ($\delta = 0.7816$, CI: [0.7552, 0.8092]). The positive values indicate that turbulent regime values systematically exceeded stable regime values, consistent with enhanced dynamical complexity.

\begin{figure}[H]
    \centering
    \includegraphics[width=\linewidth]{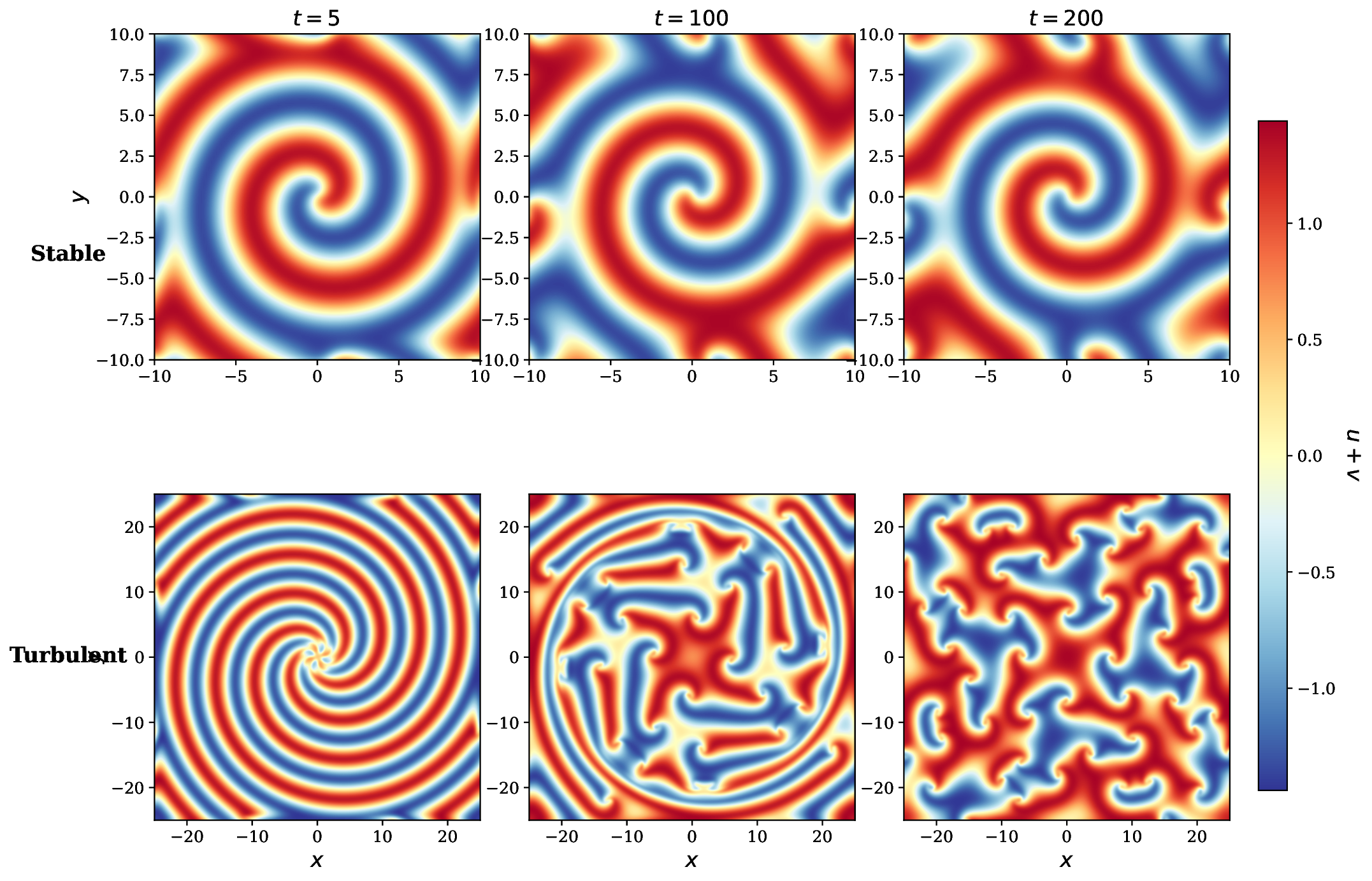}
    \caption{Spatiotemporal evolution of spiral wave patterns in stable (top row) and turbulent (bottom row) regimes. Columns show snapshots at $t = 5$, $t = 100$, and $t = 200$ time units. Color scale represents the $u + v$ concentration field ranging from -2 (blue) to +2 (red). Stable spiral maintains coherent single-armed rotation throughout the simulation, while turbulent spiral exhibits progressive fragmentation and development of spatiotemporal chaos. Domain size: $L = 20$ (stable), $L = 50$ (turbulent).}
    \label{fig:f2}
\end{figure}

The spatiotemporal patterns visualized in Figure \ref{fig:f2} provided direct confirmation of these statistical findings. Information-theoretic analysis using Shannon entropy quantified the complexity of these patterns across multiple scales. At the initial measurement time ($t = 5$), both systems showed comparable global entropy values. The stable spiral exhibited $H(u) = 6.453046$ bits, $H(v) = 6.454239$ bits, and combined field entropy $H(u+v) = 6.462929$ bits. The turbulent system showed similar values: $H(u) = 6.466953$ bits, $H(v) = 6.423104$ bits, and $H(u+v) = 6.455867$ bits.

However, spatial entropy analysis, computed as the mean entropy over $5 \times 5$ local windows, revealed important structural differences. The stable spiral showed spatial entropy of $3.668062 \pm 0.417121$ bits for the $u$-field, while the turbulent case exhibited $3.611074 \pm 0.475047$ bits, with the larger standard deviation indicating greater spatial heterogeneity. Joint entropy analysis further illuminated the coupling between concentration fields. The joint entropy $H(U,V) = 8.320609$ bits for stable spirals and $H(U,V) = 8.250349$ bits for turbulent systems, combined with individual entropies, yielded mutual information values of $I(U;V) = 4.586676$ bits (normalized: 0.710711) and $I(U;V) = 4.639707$ bits (normalized: 0.719894) respectively, indicating strong activator-inhibitor coupling in both regimes.

As the systems evolved, distinct temporal trends emerged. By $t = 100$, the stable system maintained consistent entropy levels ($H(u+v) = 6.489512$ bits, spatial entropy $3.670517 \pm 0.382467$ bits), while the turbulent system showed increased complexity ($H(u+v) = 6.600704$ bits, spatial entropy $3.660448 \pm 0.403447$ bits). Notably, the mutual information decreased in both systems but more dramatically in the turbulent case: from initial values to $I(U;V) = 4.304781$ bits (normalized: 0.663998) for stable and $I(U;V) = 3.691138$ bits (normalized: 0.564959) for turbulent dynamics.

\begin{figure}[H]
    \centering
    \includegraphics[width=\linewidth]{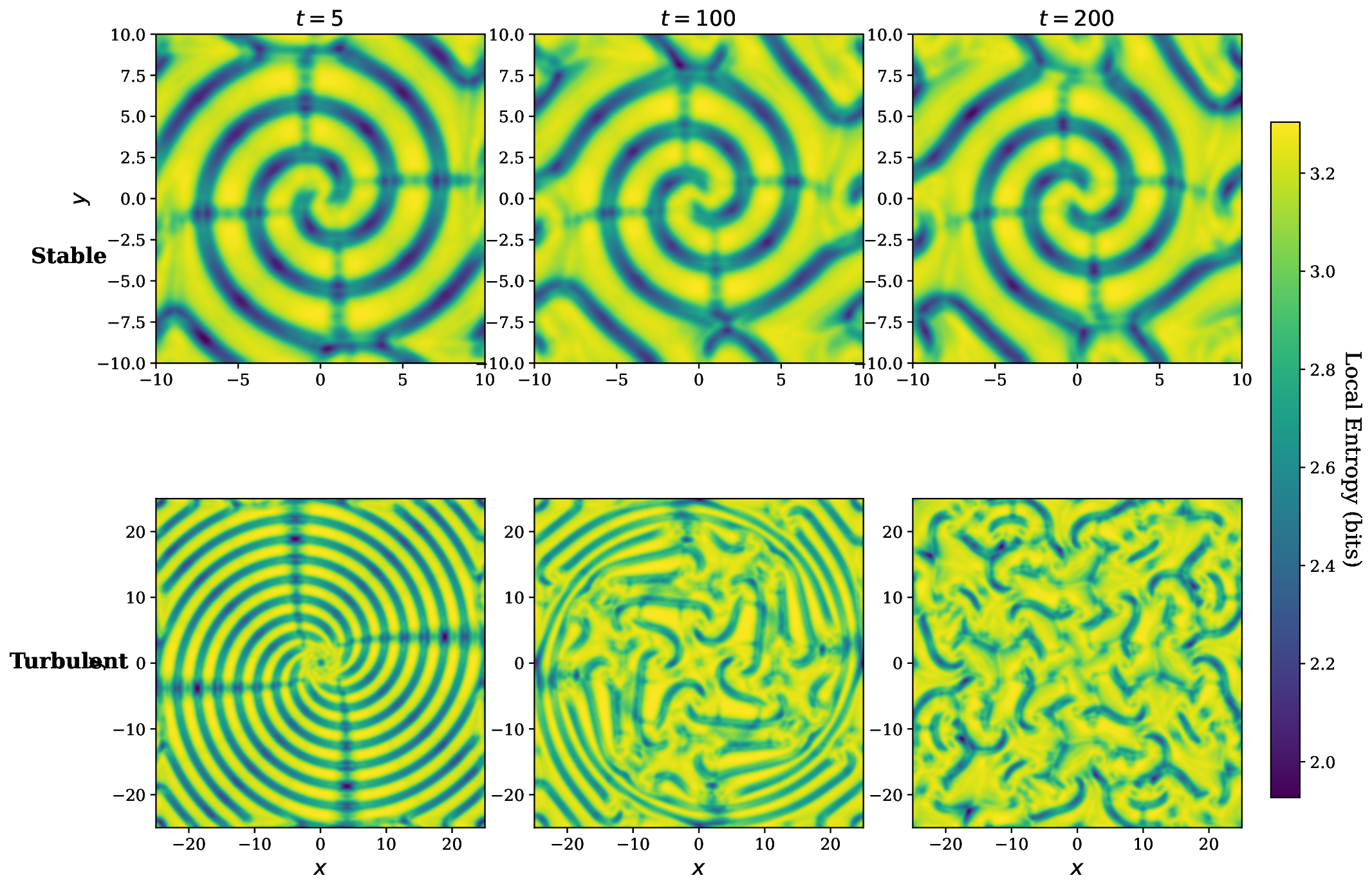}
    \caption{Spatial distribution of local Shannon entropy for stable (top row) and turbulent (bottom row) spiral waves at $t = 5$, $t = 100$, and $t = 200$. Color scale represents local entropy values computed over $5 \times 5$ windows, ranging from 2.0 bits (low complexity, dark blue) to 3.2 bits (high complexity, yellow). Stable spirals exhibit relatively uniform entropy distribution with clear spiral structure, while turbulent dynamics show heterogeneous high-entropy regions corresponding to areas of pattern fragmentation and defect formation.}
    \label{fig:f3}
\end{figure}

By the final measurement time ($t = 200$), the entropy evolution patterns had stabilized but maintained their distinctive characteristics. The stable system showed $H(u+v) = 6.487587$ bits with spatial entropy $3.682089 \pm 0.377957$ bits, while the turbulent system exhibited $H(u+v) = 6.503733$ bits with spatial entropy $3.720122 \pm 0.306543$ bits. The decreasing standard deviations of spatial entropy (from 0.417121 to 0.377957 bits for stable, and 0.475047 to 0.306543 bits for turbulent) indicated progressive spatial homogenization in both regimes, though through fundamentally different mechanisms.

Comparative analysis through entropy ratios revealed subtle but systematic differences. The turbulent-to-stable Shannon entropy ratio evolved from 0.999 at $t = 5$ to 1.017 at $t = 100$ and 1.002 at $t = 200$, while spatial entropy ratios showed similar modest variations (0.987, 0.997, and 1.010 respectively). The mutual information ratio, however, exhibited more substantial evolution, decreasing from 1.012 to 0.857 to 0.966, suggesting progressive decoupling of concentration fields in the turbulent regime.

The temporal evolution summary revealed that stable systems underwent minimal changes, with Shannon entropy increasing by only 0.4\% and spatial entropy by 0.3\% over the entire simulation period, while mutual information decreased by 6.5\%. The turbulent system showed more pronounced evolution: Shannon entropy increased by 0.7\%, spatial entropy by 2.7\%, and mutual information decreased by 10.7\%. These trends, visualized in the spatial entropy maps of Figure \ref{fig:f3}, demonstrated that stable spirals maintained relatively uniform entropy distributions preserving clear spiral structure, while turbulent dynamics generated heterogeneous high-entropy regions corresponding to pattern fragmentation and defect formation sites.

\section{Discussion}

The statistical analysis reveals fundamental dynamical distinctions between stable and turbulent spiral wave regimes that warrant careful interpretation. The extreme leptokurtosis observed in stable spirals (excess kurtosis values exceeding 96) constitutes an unprecedented departure from Gaussian behavior in spatiotemporal pattern analysis. This statistical signature indicates confinement to an exceptionally narrow region of phase space, consistent with theoretical predictions for rigidly rotating spiral waves \cite{Barkley1991}. The contrasting platykurtic distributions in turbulent regimes demonstrate exploration of a substantially broader dynamical repertoire, characteristic of systems exhibiting weak turbulence as analyzed by Cross and Hohenberg \cite{Cross1993}.

The non-parametric analysis, necessitated by these profound distributional anomalies, yielded effect sizes that illuminate differential field sensitivities. The medium effect size for $\sigma_u$ (Cliff's $\delta = 0.37$) versus the large effect for $\sigma_v$ ($\delta = 0.78$) suggests asymmetric responses to parameter perturbations. This asymmetry likely reflects the distinct dynamical roles of activator and inhibitor species, with the larger effect size for the inhibitor field consistent with theoretical predictions that diffusion coefficient disparities primarily impact the slower-diffusing component \cite{Kuramoto2003}.

Information-theoretic measures provide quantitative insight into spiral breakup mechanisms. The differential decline in mutual information—6.5\% for stable systems versus 10.7\% for turbulent dynamics—directly quantifies the progressive decoupling between concentration fields during pattern fragmentation. This metric captures the essence of defect-mediated turbulence, wherein local phase singularities disrupt the activator-inhibitor coupling required for coherent spiral maintenance \cite{Davidenko1992}. The relatively modest entropy increases (below 3\%) in both regimes indicate approach to distinct statistical steady states through fundamentally different dynamical pathways.

The pedagogical implications merit emphasis. These extreme statistical properties provide concrete demonstrations of concepts typically presented abstractly: deterministic systems generating profoundly non-Gaussian distributions, the critical importance of distribution-free statistical methods, and the physical interpretation of information-theoretic measures. The \texttt{rd-spiral} implementation enables direct exploration of these phenomena, facilitating the computational approach to nonlinear dynamics education advocated by Strogatz \cite{Strogatz2015}.

The pseudo-spectral methodology demonstrated particular advantages for this system class. The exponential convergence properties of Fourier methods \cite{Trefethen2000} proved essential for resolving steep gradients near spiral cores while maintaining numerical stability across extended temporal integration. The successful application of Dormand-Prince adaptive integration, despite substantial stiffness from disparate diffusion coefficients, validates contemporary ODE solver robustness when coupled with appropriate spatial discretization strategies \cite{Kassam2005}.

Several methodological constraints require acknowledgment. Periodic boundary conditions, while facilitating spectral accuracy, potentially suppress boundary-induced instabilities relevant to finite domains. The two-dimensional restriction excludes three-dimensional phenomena including scroll wave dynamics critical in volumetric excitable media \cite{Keener1988}. Future investigations might incorporate adaptive spatial resolution or explore emerging physics-informed neural network approaches \cite{Raissi2019} for enhanced pattern characterization.

The broader implications extend significantly beyond the specific model examined. The statistical framework developed—integrating robust hypothesis testing, effect size quantification, and information-theoretic analysis—establishes a rigorous template for quantitative pattern characterization across diverse dynamical systems. The extreme distributional properties documented here suggest similar departures from normality likely pervade other deterministic pattern-forming systems, necessitating appropriate statistical treatment frequently absent in current literature. These findings bear immediate relevance for cardiac arrhythmia dynamics \cite{Winfree1987}, intracellular calcium wave propagation \cite{Lechleiter1991}, and ecological pattern formation \cite{Hassell1991}, where analogous spiral phenomena occur but comprehensive statistical characterization remains incomplete.

\section{Conclusions}

The development of \texttt{rd-spiral} represents a significant contribution to computational methods for reaction-diffusion systems, demonstrating that pedagogically-oriented software can achieve research-quality results while maintaining algorithmic transparency. Our pseudo-spectral implementation successfully captured the full spectrum of spiral wave dynamics—from stable rotation to spatiotemporal chaos—revealing extreme statistical properties (excess kurtosis $>$ 96) that necessitate robust non-parametric analysis. The solver's architecture, combining FFT-based spatial discretization with adaptive time integration, proved remarkably efficient across disparate parameter regimes despite stiffness ratios exceeding 6:1. Beyond the statistical framework establishing quantitative distinctions between dynamical regimes (Cliff's $\delta = 0.37–0.78$, mutual information decline 6.5–10.7\%), the open-source implementation provides a validated computational platform that bridges theoretical understanding with practical exploration. By prioritizing code clarity, comprehensive documentation, and reproducible parameter sets, \texttt{rd-spiral} addresses the critical need for accessible yet rigorous tools in nonlinear dynamics education while establishing computational protocols applicable to broader classes of pattern-forming systems where traditional numerical approaches often obscure the underlying mathematical elegance.

\section*{Acknowledgements}
This study was supported by the Dean's Distinguished Fellowship from the College of Natural and Agricultural Sciences (CNAS) at the University of California, Riverside in 2023 and ITB Research, Community Service and Innovation Program (PPMI-ITB) in 2025.

\section*{Open Research}

The \texttt{rd-spiral} library source code is available at \url{https://github.com/sandyherho/rd_spiral} under the MIT license. Python code for statistical analyses is available at \url{https://github.com/sandyherho/data_analysis_rd-spiral}. All simulation outputs are archived at \url{https://doi.org/10.17605/OSF.IO/UYGVQ} (Open Science Framework) under the MIT license.

\section*{Author Contributions}
\textbf{S.H.S.H.}: Conceptualization; Formal analysis; Methodology; Software; Visualization; Writing – original draft. \textbf{I.P.A.}: Supervision; Writing – review \& editing. \textbf{R.S.}: Supervision; Writing – review \& editing. All authors reviewed and approved the final version of the manuscript.

\end{document}